\documentclass[conference]{IEEEtran}

\usepackage{hyperref}
\usepackage{mathtools}
\usepackage{epsfig,amsmath,amssymb,amsthm}
\usepackage{xcolor}
\newtheorem{theorem}{Theorem}
\usepackage{mdframed}   
\usepackage{comment}

\usepackage{array}
\usepackage{booktabs}
\usepackage{multirow}

\newmdtheoremenv[%
linecolor=gray,leftmargin=0,%
rightmargin=0,
backgroundcolor=gray!20,%
innertopmargin=0pt,%
ntheorem]{myprop}{Our take-away}[section]

\usepackage{algorithm}
\usepackage{algpseudocode}

\newcommand{\subsubsec}[1] {\vspace*{0.075in}\noindent\textbf{#1:}}
\newcommand{\subsubsecb}[1] {\noindent\textbf{#1:}}
\newcommand{\boldpara}[1] {\vspace*{0.075in}\noindent\underline{\emph{#1:}}}

\providecommand{\abs}[1]{\lvert#1\rvert}

\begin{document}

\title{SoK: Towards the Science of \\ Security and Privacy in Machine Learning}
\author{Nicolas Papernot\IEEEauthorrefmark{1}, Patrick McDaniel\IEEEauthorrefmark{1}, Arunesh Sinha\IEEEauthorrefmark{2}, and Michael Wellman\IEEEauthorrefmark{2} \\ \IEEEauthorrefmark{1} Pennsylvania State University \\ \IEEEauthorrefmark{2} University of Michigan \\ \normalsize{\texttt{\{ngp5056,mcdaniel\}@cse.psu.edu, \{arunesh,wellman\}@umich.edu}}}

\maketitle

\thispagestyle{plain}
\pagestyle{plain}

\begin{abstract}
Advances in machine learning (ML) in recent years have enabled a dizzying array of applications such as data analytics, autonomous systems, and security diagnostics.  ML is now pervasive---new systems and models are being deployed in every domain imaginable, leading to rapid and widespread deployment of software based inference and decision making. There is growing recognition that ML exposes new vulnerabilities in software systems, yet the technical community's understanding of the nature and extent of these vulnerabilities remains limited. We systematize recent findings on ML security and privacy, focusing on attacks identified on these systems and defenses crafted to date.  We articulate a comprehensive threat model for ML, and categorize attacks and defenses within an adversarial framework.  Key insights resulting from works both in the ML and security communities are identified and the effectiveness of approaches are related to structural elements of ML algorithms and the data used to train them. We conclude by formally exploring the opposing relationship between model accuracy and resilience to adversarial manipulation.  Through these explorations, we show that there are (possibly unavoidable) tensions between model complexity, accuracy, and resilience that must be calibrated for the environments in which they will be used.
\end{abstract}
\IEEEpeerreviewmaketitle

\section{Introduction}
\label{sec:introduction}

The coming of age of the science of machine learning (ML) coupled with advances
in computational and storage capacities have transformed the technology
landscape.  For example, ML-driven data analytics have fundamentally altered the
practice of health care and financial management. Within the security domain,
detection and monitoring systems now consume massive amounts of data and extract
actionable information that in the past would have been impossible. Yet, in
spite of these spectacular advances, the technical community's understanding of
the vulnerabilities inherent to the design of systems built on ML and the means
to defend against them are still in its infancy. There is a broad and pressing
call to advance a science of the security and privacy in ML.

Such calls have not gone unheeded.  A number of activities have been launched to
understand the threats, attacks and defenses of systems built on machine
learning.  However, work in this area is fragmented across several research
communities including machine learning, security, statistics, and theory of
computation, and there has been few efforts to develop a unified lexicon or
science spanning these disciplines.   This fragmentation presents both a
motivation and challenge for our effort to systematize knowledge about the
myriad of security and privacy issues that involve ML.  In this paper we develop
a unified perspective on this field.  We introduce a unified threat model that
considers the attack surface and adversarial goals and capabilities of systems
built on machine learning.  The security model serves as a roadmap in the
following sections for exploring attacks and defenses of ML systems.   We draw
major themes and highlight results in the form of take-away messages about this
new area of research.  We conclude by providing a theorem of the ``no free lunch'' properties of many ML systems---identifying where there is a tension between complexity and resilience to adversarial manipulation, how this tension impacts the accuracy of models and the effect of the size of datasets on this trade-off.

In exploring security and privacy in this domain, it is instructive to view
systems built on machine learning through the prism of the classical
confidentiality, integrity, and availability (CIA) model.  In this work,
confidentiality is defined with respect to the model or its training data. 
Attacks on confidentiality attempt to expose the model structure or parameters
(which may be highly valuable intellectual property) or the data used to train
it, e.g., patient data.  The latter class of attacks have a potential to impact
the privacy of the data source, e.g., the privacy of patient clinical data used
to train medical diagnostic models is often of paramount importance. Conversely,
we define attacks on the integrity as those that induce particular outputs or
behaviors of the adversary's choosing.  Where those adversarial behaviors
attempt to prevent access to meaningful model outputs or the features of the
system itself, such attacks fall within the realm of availability.

A second perspective in evaluating security and privacy focuses on attacks and
defenses with respect to the {\it machine learning pipeline}.  Here, we consider
the lifecycle of a ML-based system from training to inference, and identify the
adversarial goals and means at each phase.  We observe that attacks on training
generally attempt to influence the model by altering  or injecting  training
samples--in essence guiding the learning process towards a vulnerable model. 
Attacks at inference time (runtime) are more diverse.  Adversaries use
exploratory attacks to induce targeted outputs, and oracle attacks to extract
the model itself.

The science of defenses for machine learning are somewhat less well developed.  
Here we consider several defensive goals.  First, we consider methods at
training and inference time that are robust to distribution drifts--the property
that ensures that the model performs adequately when the training and runtime
input distributions differ.  Second, we explore models that provide formal
privacy preserving guarantees--the property that the amount of data exposed by
the model is bounded by a privacy budget (expressed in terms of differential
privacy).  Lastly, we explore defenses that provide fairness (preventing biased
outputs) and accountability (explanations of why particular outputs were
generated, also known as transparency).

In exploring these facets of machine learning attacks and defense, we make the
following contributions:
\begin{itemize}
	
	\item We introduce a unifying threat model to allow structured reasoning about the security and privacy of systems that incorporate machine learning. 	This model, presented in  Section~\ref{sec:secmodel}, departs from previous efforts by considering the entire data pipeline, of which ML is a component, instead of ML algorithms in isolation. 
	
	\item We taxonomize attacks and defenses identified by the various technical communities as informed elements of PAC learning theory.  Section~\ref{sec:training} details the challenges of  in adversarial settings and Section~\ref{sec:inference} considers trained and deployed systems. Section~\ref{sec:future} presents desirable properties to improve the  security and privacy of ML.
	
	\item In Section~\ref{sec:learning-theory}, we introduce a \emph{no free lunch} theorem for adversarial machine learning. It characterizes the tradeoff between accuracy and robustness to adversarial efforts, when learning from limited data.
	
\end{itemize}

\noindent
Note that ML systems address a great many different problem domains, e.g.,
classification, regression and policy learning.  However, for brevity and ease
of exposition, we focus much of the current paper on ML classification.  We
further state that the related study of the implications of ML and AI on safety
in societies is outside the scope of this paper, and refer interested readers to
the comprehensive review by Amodei et al.~\cite{amodei2016concrete}.

The remainder of this paper focuses on a systematization of the knowledge of
security and privacy in ML. While there is an enormous body of work in many
aspects of this research, for space we focus on attacks and defenses. As a
result of a careful analysis of the threat model, we have selected seminal and
representative works that illustrate the branches of this research. While we
attempt to be exhaustive, it is a practical impossibility to cite all works. For
instance, we do not cover trusted computing platforms for
ML~\cite{ohrimenko2016oblivious}. We begin below by introducing the basic
structure and lexicon of ML systems.

\section{About Machine Learning}
\label{sec:background}

We start with a brief overview of how systems apply ML
algorithms. In particular, we compare different
kinds of learning tasks, and some specifics of their practical implementation.

\subsection{An Overview of Machine Learning Tasks}

\emph{Machine learning} provides automated methods of analysis for large sets of
data~\cite{murphy2012machine}. Tasks solved with machine learning
techniques are commonly divided into three types. These are characterized by the
structure of the data analyzed by the corresponding learning algorithm.

\boldpara{Supervised learning} Methods that induce an association between
inputs and outputs based on training examples in the form of inputs labeled with corresponding outputs are \emph{supervised learning} techniques. 
If the output data is categorical, the task is called \emph{classification}, and
real-valued output domains define \emph{regression} problems. 
Classic examples
of supervised learning tasks include: object recognition in
images~\cite{krizhevsky2012imagenet}, machine
translation~\cite{sutskever2014sequence}, and spam
filtering~\cite{drucker1999support}.

\boldpara{Unsupervised learning} When the method is given unlabeled inputs, its
task is \emph{unsupervised}. 
Unsupervised learning considers problems
such as \emph{clustering} points according to a similarity
metric~\cite{jain1999data}, \emph{dimensionality reduction} to project data in
lower dimensional subspaces~\cite{krizhevsky2009learning}, and
\emph{model pre-training}~\cite{erhan2010does}. For instance, clustering may be applied to anomaly detection~\cite{chandola2009anomaly}.

\boldpara{Reinforcement learning} Methods that learn a policy for action over time given sequences of actions, observations, and rewards  fall in the scope of \emph{reinforcement
	learning}~\cite{Hu03,sutton1998reinforcement}. 
Reinforcement learning can be viewed as the subfield of ML concerned with planning and control.
It was reinforcement learning in combination with supervised and unsupervised methods that recently enabled a computer to defeat a human
champion at the game of Go~\cite{silver2016mastering}.

\vspace*{0.08in}

Readers interested in a broad survey of ML are well served by
many books covering this rich
topic~\cite{murphy2012machine,bishop2006pattern,goodfellow2016book}. 
Work on ML security and privacy to date has for the most part conducted in supervised
settings, especially in the context of classification tasks, as reflected by our presentation in
Sections \ref{sec:training} and~\ref{sec:inference} below. 
Since security issues are just as relevant for unsupervised and reinforcement learning tasks, we strive to present results in
more general settings when meaningful.

\subsection{Data Collection: Three Use Cases}

Before one can learn a model that solves a task of interest, training
	data must be collected. This consists in gathering a generally large set of examples
of solutions to the task that one wants to solve with machine learning. 
For each of the task types introduced 
above, we describe one example of a task and how the corresponding training dataset would be collected.

The first example task is to classify software executables in two categories:
malicious and benign. This is a supervised
classification problem, where the model must learn some mapping between inputs (software executables) and
categorical outputs (this binary task only has two possible classes).
The training data comprises a set of labeled instances, each an executable clearly marked as malicious or
benign~\cite{christodorescu2006static}.

Second, consider the task of extracting a pattern representative of normal activity
in a computer network.
The training data could consist of TCP
dumps~\cite{zhang2006anomaly}. Such a scenario is commonly encountered in
anomaly-based network intrusion detection~\cite{sommer2010outside}. Since the
model's desired outputs are not given along with the input---that is, the TCP dumps
are not associated with any pattern specification---the
problem falls under the scope of unsupervised learning. 

Finally, consider the
same intrusion-detection problem given access to metrics of system state
indicator (CPU load, free memory, network load, etc.)~\cite{cannady2000next}.
This variant of the intrusion detector can then be viewed as an agent and the
system state indicator as rewards for actions taken based on a prediction 
made by the intrusion
detector (e.g., shut down part of the network infrastructure). In this form,
the scenario then falls under the reinforcement learning tasks.

\subsection{Machine Learning Empirical Process} 

We describe the general approach taken to create a machine learning model
solving one of the tasks described above. 

\boldpara{Training} Once the data is collected and
pre-processed, a ML model is chosen and trained. Most\footnote{A few models are non-parametric: for instance the nearest neighbor~\cite{altman1992introduction}.} ML models can be seen as parametric functions
$h_\theta(x)$ taking an input $x$ and a parameter vector $\theta$. 
The input $x$ is often represented as a vector of values called \emph{features}.
The space of functions $\{\forall \theta, x\mapsto h_\theta(x)\}$ is the set of candidate hypotheses to model the distribution from which the dataset was sampled. A \emph{learning
algorithm} analyzes the training data to find the value(s) of parameter(s) $\theta$. When learning is supervised, the parameters
are adjusted to reduce the gap between model predictions $h_\theta(x)$ and the expected
output indicated by the dataset. In reinforcement learning, the agent
adjusts its policy to take actions that yield the highest reward. The
model performance is then validated on a test dataset, which must be
disjoint from the training dataset in order to measure the model's \emph{generalization}.
For a supervised problem like malware classification
(see above), the learner computes the model \emph{accuracy}
on a test dataset, i.e. the proportion of predictions
$h_\theta(x)$ that matched the label $y$ (malware or benign) associated with the
executable $x$ in the dataset. 
When learning is done in an online fashion, parameters $\theta$ are updated as new 
training points become available.

\boldpara{Inference} Once training completes, the model is deployed to
\emph{infer} predictions on inputs unseen during training: i.e., the value of  
parameters $\theta$ are fixed, and the model
computes $h_\theta(x)$ for new inputs $x$. In our running example,
the model would predict whether an executable $x$ is more likely to be
malware or benign. The model prediction may take different forms but the most common
for classification is a vector assigning a probability for each class of
the problem, which characterizes how likely the input is to belong to that class. 
For our unsupervised network intrusion detection system, 
the model would instead return the pattern representation $h_\theta(x)$ 
that corresponds to a new input network traffic $x$.

\subsection{A Theoretical Model of Learning}
Next, we formalize the semantics of supervised ML algorithms. We give an
overview of the Probably Approximately Correct (PAC) model, a theoretical underpinning of these algorithms, here and later use the model in
Sections~\ref{sec:training},~\ref{sec:inference}, and~\ref{sec:future} to
interpret attacks and defenses. Such an interpretation helps discover, from
specific attacks and defenses, general principles of adversarial learning that
apply across all supervised ML.

\boldpara{PAC model of learning} PAC learning model has a very rich and
extensive body of work~\cite{anthony2009neural}. Briefly, the PAC model states that data points
$(x,y)$  are samples obtained by sampling from a fixed but unknown probability
distribution $D$ over the space $Z = X \times Y$. Here, $X$ is the space of
feature values and $Y$ is the space of labels (e.g., $Y= \{0,1\}$ for
classification or $Y=\mathbb{R}$ for regression). The mapping from $X$ to
$Y$ is captured by a function $h:X \rightarrow Y$ 
associated with a loss function $l_h:X \times Y \rightarrow \mathbb{R}$ which
captures the error made by the prediction $h(x)$ when the true label is $y$.
Examples include the hinge loss~\cite{rosasco2004loss} used in SVMs 
or the cross-entropy loss~\cite{murphy2012machine}. 
The learner
aims to learn a function $h^*$ from a family of functions $\mathcal{H}$ such
that the the expected loss (also called risk) $r(h) = E_{x,y \sim D}[l_h(x,y)]$
is minimal, that is,
$
h^* \in \arg\!\min_{h\in \mathcal{H}} r(h)
$.

Of course, in practice $D$ is not known and only samples $\vec{z} = z_1, \ldots, z_n$ (the training data) is observed. The learning algorithm then uses the empirical loss $\hat{r}(h, \vec{z}) = \frac{1}{n}\sum_{i=1}^n l_h(x_i,y_i)$, where $z_i=(x_i,y_i)$ as a proxy for the expected loss and finds the $h$ that is close to 
$
\hat{h} \in \arg\!\min_{h\in \mathcal{H}} \hat{r}(h, \vec{z})
$. Thus, all supervised learning algorithm perform this \emph{empirical risk minimization} (ERM) with the loss function varying across different algorithms.
The PAC guarantee states that:
\begin{equation}
P(\abs{r(h^*) - r(\hat{h})} \leq \epsilon) \geq 1 - \delta
\end{equation}
where the probability is over samples $\vec{z}$ used to learn $\hat{h}$. This guarantee holds when 
two pre-conditions are met: \textbf{[Condition 1: Uniform bound]} given enough samples (called the sample complexity, which depends on $\epsilon,\delta$ above) that enable a uniform bound of the difference between the actual and empirical risk for all functions in $\mathcal{H}$, and \textbf{[Condition 2: Good ERM]} $\hat{h}$ is close to the true empirical risk minimizer $\hat{h}$.

Statistical learning is primarily concerned with the uniform bound pre-condition above, wherein a good ERM is assumed to exist and the goal is to find the sample complexity required to learn certain classes of function with certain loss function.

The training step in supervised learning algorithms performs the ERM step.
The accuracy measured on the test data in machine learning typically estimates the $\epsilon$ (or some error value correlated with $\epsilon$). In particular, the train test procedure relies on the assumption that training and test data arise from the same, though unknown, distribution $D$ and more importantly the distribution faced in an actual deployment in the inference step is also $D$. Later we show that most attacks arise from an adversarial modification of $D$ either in training or in inference resulting in a mismatch between the distribution of data used in the learning and the inference phases.

Another noteworthy point is that the PAC guarantee (and thus most ML algorithms) is only about the expected loss. Thus, for most data points $(x,y)$ that lie in low probability regions, the predictor $\hat{h}(x)$ can be far from the true $y$ yet the output $\hat{h}$ could have high accuracy (as measured by test accuracy) because the accuracy is an estimate of the expected loss. In the extreme, a learning accuracy of 100\% could be achieved by predicting correctly in the positive probability region with lot of misclassification in the zero probability regions (or more precisely sets with measure zero; a detailed discussion of this fact is present in a recent paper~\cite{sinha2016learning}). An adversary may exploit such misclassification to its advantage. We elaborate on the two points above later in Section~\ref{sec:learning-theory}.

\section{Threat Model}
\label{sec:secmodel}
\label{sec:ml-security}

The security of any system is measured with respect the adversarial goals and capabilities that it is designed to defend against--the systems' {\it threat model}.  In this section we taxonomize the definition and scope of threat models in machine learning systems and map the space of security models.  We begin by identifying the threat surface of systems built on machine learning to inform where and how an adversary will attempt to subvert the system under attack. 
For the purpose of exposition of the following Sections, we expand upon previous approaches at articulating a threat model for ML~\cite{barreno2006can,papernot2015limitations}.

\begin{figure*}
	\centering
	\includegraphics[width=0.9\textwidth]{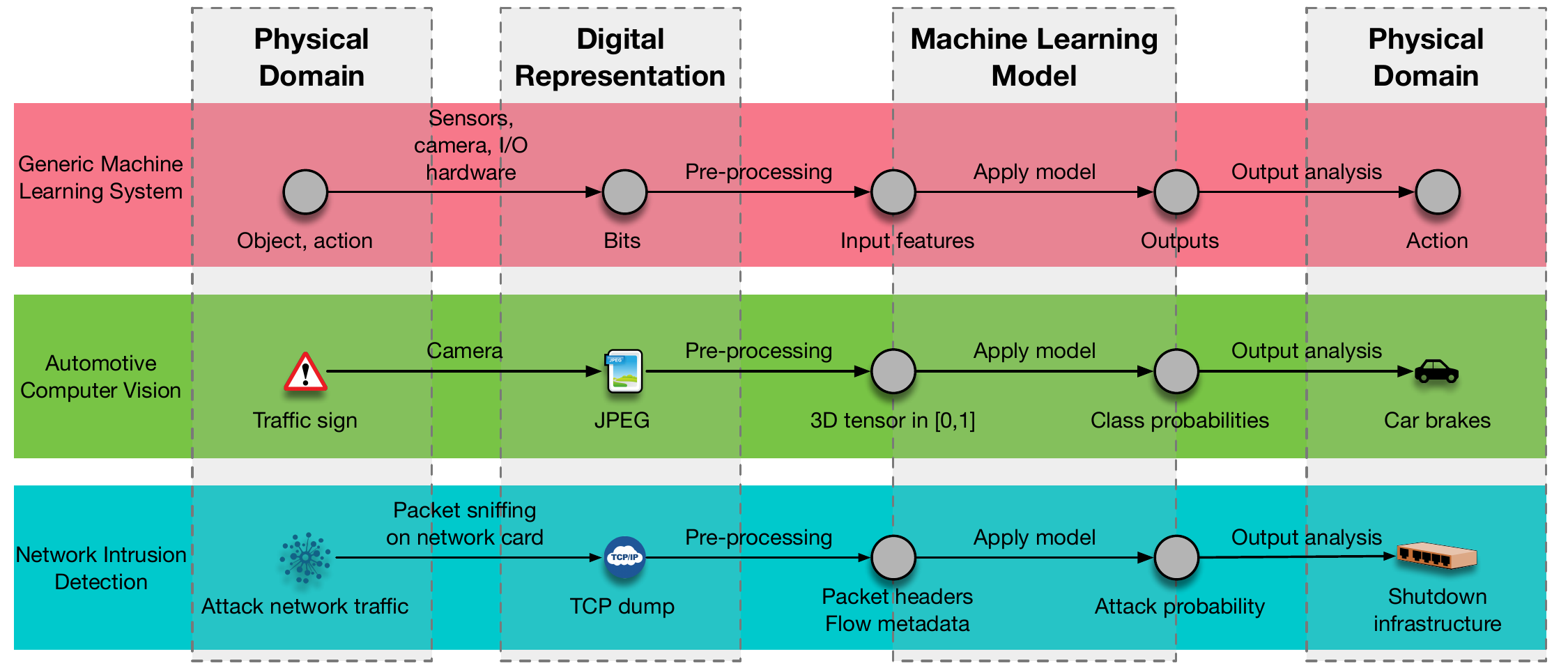}
	\caption{\textbf{System's attack surface:} the
		generic model (top row) is illustrated with two example scenarios (bottom rows):
		a computer vision model used by an automotive system to recognize traffic signs 
		on the road and a network intrusion detection system.}
	\label{fig:ml-attack-surface}
\end{figure*}

\subsection{The ML Attack Surface}
\label{ssec:mlpipe}

The attack surface of a system built with data and machine learning is reflective of its purpose.  However, one can view systems using ML within a generalized data processing pipeline (see Figure~\ref{fig:ml-attack-surface}, top).  At inference, (a) input features are collected from sensors or data repositories, (b) processed in the digital domain, (c) used by the model to produce an output, and (d) the output is communicated to an external system or user and  acted upon.  To illustrate, consider a generic pipeline, autonomous vehicle, and network intrusion detection systems in Figure~\ref{fig:ml-attack-surface} (middle and bottom).  These systems collect sensor inputs (video image, network events) from which model features (pixels, flows) are extracted and used within the models.  The meaning of the model output (stop sign, network attack) is then interpreted and action taken (stopping the car, filtering future traffic from an IP). Here, the attack surface for the system can be defined with respect to the data processing pipeline.  Adversaries can attempt to manipulate the collection and processing of data, corrupt the model, or tamper with the outputs.

Recall that the training of the model is performed using either an offline or online  process.  In an offline setting, training data is collected or generated.  The training data used to learn the model includes vectors of features used as inputs during inference, as well as expected outputs for supervised learning or a reward function for reinforcement learning.  The training data may also include additional features not available at runtime (referred to as privileged information in some settings~\cite{vapnik2009new}).  As discussed below, the means of collection and validation processes offer another attack surface--adversaries who can manipulate the data collection process can do so to induce targeted model behaviors.  Similar attacks in an online setting (such as may be encountered in reinforcement learning) can be quite damaging, where the adversary can slowly alter the model with crafted inputs submitted at runtime (e.g., using false training~\cite{kloft2010online}). Online attacks such as these have been commonly observed in domains such as SPAM detection and network intrusion detection~\cite{kloft2010online}.

\subsection{Adversarial Capabilities}
\label{ssec:adcaps}

\begin{figure}
	\centering
	\includegraphics[width=0.4\textwidth]{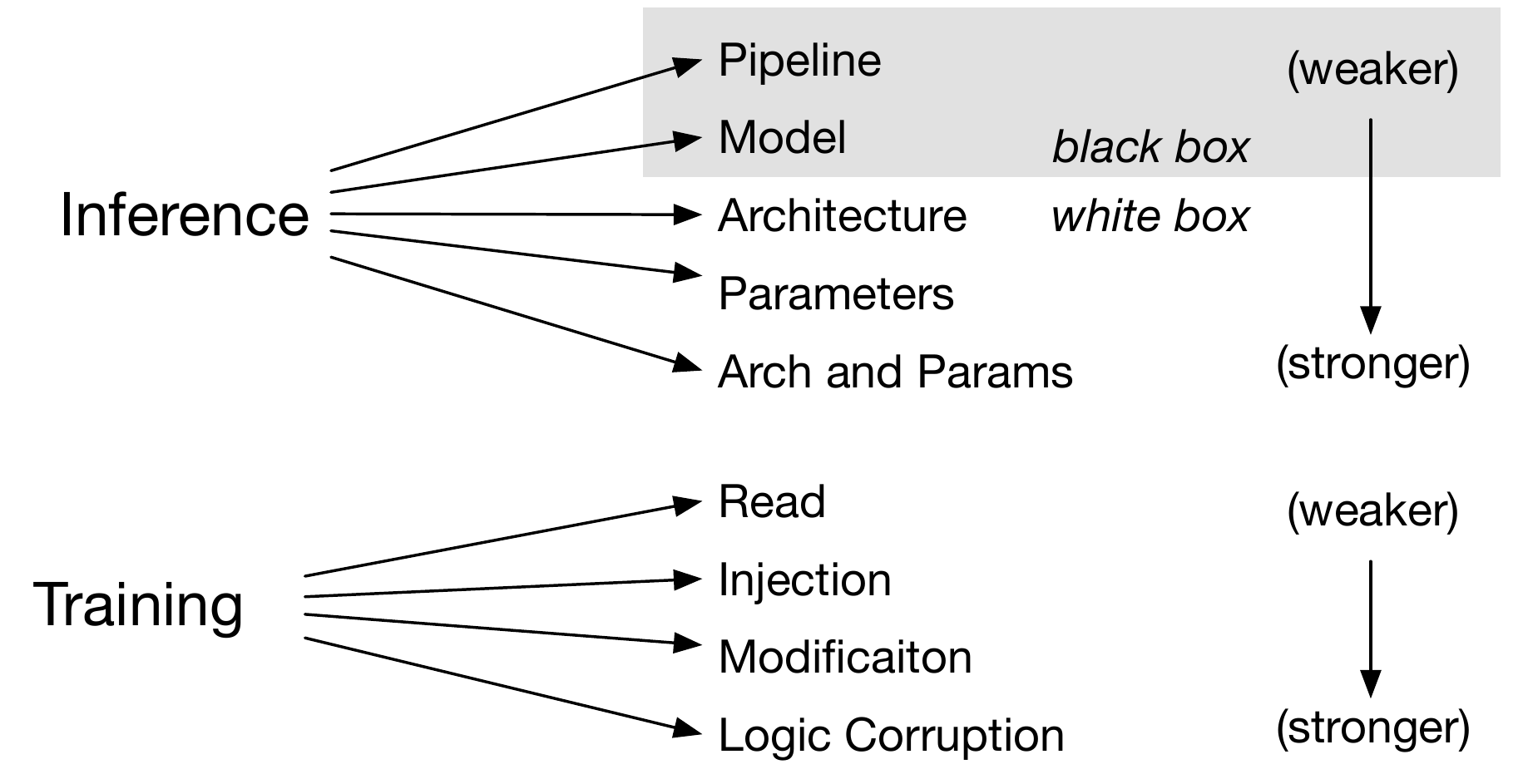}
	\caption{\textbf{Adversarial Capabilities:} adversaries attack ML systems at inference time by exploiting model internal information (white box) or probing the system to infer system vulnerabilities (black box).  Adversaries use read or write access to the training data to mimic or corrupt the model. }
	\label{fig:ml-capabilities}
\end{figure}

A threat model is also defined by the actions and information the adversary has at their disposal.  The definition of security is made with respect to stronger or weaker adversaries who have more or less access to the system and its data.  The term capabilities refers to the whats and hows of the available attacks, and indicates the attack vectors one may use on a threat surface.   For instance, in the network intrusion detection scenario, an internal adversary may have access to the model used to distinguish attacks from normal behavior, whereas a weaker eaves-dropping adversary would only have access to TCP dumps of the network traffic. Here the attack surface remains the same for both the attacks, but the former attacker is assumed to have much more information and is thus a strictly "stronger" adversary.   We explore the range of attacker capabilities in machine learning systems as they relate to inference and training phases (see Figure~\ref{fig:ml-capabilities}).

\boldpara{Inference Phase}
Attacks at inference time---\emph{exploratory attacks}~\cite{barreno2006can}---do not tamper with the targeted model but instead either cause it to produce adversary selected outputs (incorrect outputs, see Integrity attacks below) or simply use the attack to collect evidence about the model characteristics (reconnaissance, see privacy below).   As explored at length in Section~\ref{sec:inference}, the effectiveness of such attacks are largely determined by the amount of information that is available to the adversary about the model and its use in the target environment.  

Inference phase attacks can be classified into either white box or black box attacks.  In white box attacks, the adversary has some information about the model or its original training data, e.g., ML algorithm $h$, model parameters $\theta$, network structure, or summary, partial, or full training data.  Grossly speaking, this information can be divided into attacks that use information about the model architecture (algorithm and structure of the hypothesis $h$), model parameters $\theta$ (weights), or both.  The adversary exploits available information to identify where a model may be exploited.  For example, an adversary who has access to the model $h$ and its parameters $\theta$ may identify parts of the feature space for which the model has high error, and exploit that by altering an input to into that space, e.g., adversarial example crafting~\cite{szegedy2013intriguing}.

Conversely black box attacks assume no knowledge about the model.  The adversary in these attacks use information about the setting or past inputs to infer model vulnerability.  For example, in a oracle attack, the adversary explores a model by providing a series of carefully crafted inputs and observing outputs~\cite{papernot2016practical}.  Oracle attacks work because a good deal of information about a model can be extracted from input / output pairs, and relatively little information is required because of the transferability property exhibited by many model architectures (See Section~\ref{ssec:inference-blackbox}).

\boldpara{Training Phase} 
Attacks on training attempt to learn, influence or corrupt the model itself.  The simplest and arguably weakest attack on training is simply accessing a summary, partial or all of the training data.  Here, depending on the quality and volume of training data, the adversary can create a substitute model (also referred to as a surrogate or auxiliary model) to use to mount attacks on the victim system.  For example, the adversary can use a substitute model to test potential inputs before submitting them to the victim system~\cite{laskov2014practical}.  Note that these attacks are offline attempts at model reconnaissance, and thus may be used to undermine privacy (see below). 

There are two broad attack strategies for altering the model.  The first alters the training data  either by inserting adversarial inputs into the existing training data (injection) or altering the training data directly (modification).  In the case of reinforcement learning, the adversary may modify the environment in which the agent is evolving.  Lastly, the adversaries can tamper with the learning algorithm.  We refer to these attacks as logic corruption.  Obviously, any adversary that can alter the learning logic (and thus controls the model itself) is very powerful and difficult to defend against.

\subsection{Adversarial Goals}
\label{ssec:adgoals}

The last piece of a threat model is an articulation of the goals of the adversary.  We adopt a classical approach to modeling adversarial goals by modeling desired ends as impacting confidentiality, integrity, and availability (called a CIA model), and adding a fourth property, privacy.  Interestingly, a duality emerges when taking a view in this way:  attacks on system integrity and availability are closely related in goal and method, as are confidentiality and privacy.

As is often the case in security, ML systems face three types of risks: failure to provide integrity, availability, and privacy.  Integrity and privacy can both be understood at the level of the ML model itself, as well as for the entire system deploying it. Availability is however ill defined for a model in isolation but makes sense for the ML-based system as a whole. We discuss below the range of adversarial goals that relate to each risk.

\boldpara{Confidentiality and Privacy} 
Attacks on confidentiality and privacy are with respect to the model.  Put another way, the attacks achieving these goals attempt to extract information about the model or training data as highlighted above.  When the model itself  represents intellectual property, it requires that the model and its parameters be confidential, e.g., financial market systems~\cite{bolton2002statistical}.  In other contexts it is imperative that the privacy of the  training data be preserved, e.g., medical applications~\cite{rindfleisch1997privacy}.   Regardless of the goal, the attacks and defenses for them relate to exposing or preventing the exposure of the model and training data.

Machine learning models have enough capacity to capture and memorize elements of their training data~\cite{fredrikson2015model}. As such, it is hard to provide guarantees that participation in a dataset does not harm the privacy of an individual. Potential risks are adversaries performing membership test (to know whether an individual is in a dataset or not)~\cite{shokri2016membership}, recovering of partially known inputs (use the model to complete an input vector with the most likely missing bits), and extraction of the training data using the model's predictions~\cite{fredrikson2015model}.

\boldpara{Integrity and Availability} 
Attacks on integrity and ability are with respect to model outputs.   Here the goal is to induce model behavior as chosen by the adversary.  Attacks attempting to control model outputs are at the heart of integrity attacks---the integrity of the inference process is undermined.   For example, attacks that attempt to induce false positives in a face recognition system affect the authentication process's integrity~\cite{sharif2016accessorize}.  Closely related, attacks on availability attempt to reduce the quality (e.g., confidence or consistency), performance (e.g., speed), or access (e.g., denial of service).  Here again, while the goals of these two classes of attacks may be different, the means by which the adversary achieves them is often similar.

Integrity is essential in ML, and is the center of attention for most performance metrics used: e.g., accuracy~\cite{powers2011evaluation}. However, researchers have shown that the integrity of ML systems may be compromised by adversaries capable of manipulating model inputs~\cite{szegedy2013intriguing} or its training data~\cite{kearns1993learning}.  First, the ML model's confidence may be targeted by an adversary: reducing this value may change the behavior of the overall system. For instance, an intrusion detection system may only raise an alarm when its confidence is over a specified threshold. Input misprocessing aims at misleading the model into producing wrong outputs for some inputs, either modified at the entrance of the pipeline, or at the input of the model directly. Depending on the task type, the wrong outputs differ. For a ML classifier, it may assign the wrong class to a legitimate image, or classify noise with confidence. For an unsupervised feature extractor, it may produce a meaningless representation of the input. For a reinforcement learning agent, it may act unintelligently given the environment state. However, when the adversary is capable of subverting the input-output mapping completely, it can control the model and the system's behavior. For instance, it may force an automotive's computer vision system to misprocess a traffic sign, resulting in the car accelerating.

Availability is somewhat different than integrity, as it is about the prevention of access to an asset--the asset being an output or an action induced by a model output.  Hence, the goal of these attacks is to make the model inconsistent or unreliable in the target environment.  For example, the goal of the adversary attacking an autonomous vehicle may be to get it to behave erratically or non-deterministically in a given environment.  Yet most of the attacks in this space require corrupting the model through training input poisoning and other confidence reduction attacks using many of the same methods used for integrity attacks.  

If the system depends on the output of the ML model to take decisions that impact its availability, it may be subject to attacks falling under the broad category of denial of service. Continuing with the previous example, an attack that produces vision inputs that force a autonomous vehicle to stop immediately may cause a denial of service by completely stopping traffic on the highway.   More broadly, machine learning models may also not perform correctly when some of their input features are corrupted or missing~\cite{globerson2006nightmare}.  Thus, by denying access to these features we can subvert the system.

\section{Training in Adversarial Settings}
\label{sec:training}

As parameters $\theta$ of the hypothesis $h$ are fine-tuned during learning, the
training dataset analyzed is potentially vulnerable to manipulations by
adversaries. This scenario corresponds to a \emph{poisoning
	attack}~\cite{barreno2006can}, and is an instance of learning in the presence of
non-necessarily adversarial but nevertheless noisy data~\cite{manwani2013noise}.
Poisoning attacks alter the training dataset by inserting, editing, or removing
points with the intent of modifying the decision boundaries of the targeted
model~\cite{kearns1993learning}, thus targeting the learning system's integrity
per our threat model from Section~\ref{sec:secmodel}. It is somewhat obvious
that an unbounded adversary can cause the learner to learn any arbitrary
function $h$ leading to complete unavailability of the system. Thus, all the
attacks below bound the adversary in their attacks~\cite{nelson2006bounding}.
Also, in the PAC model, modifications of the training data can be seen as
altering the distribution $D$ that generated the training data, thereby creating
a mismatch between the distributions used for training and inference. In
Section~\ref{ssec:future-robust}, we present a line of work that builds on that
observation to propose learning strategies robust to \emph{distribution
drifts}~\cite{hulten2001mining}.

Upon surveying the field, we note that works almost exclusively discuss
poisoning attacks against classifiers (supervised models trained with
labeled data). However, as we strive to generalize our observations to other
types of machine learning tasks (see Section~\ref{sec:background}), we note that
the strategies described below may apply, as for instance many 
algorithms used for reinforcement learning use supervised submodels.

\subsection{Targeting Integrity}

Kearns et al.  formally analyzed PAC-learnability when the adversary is allowed to modify training samples
with probability $\beta$~\cite{kearns1993learning}. For large datasets this adversarial capability can be interpreted as the ability to modify a  fraction $\beta$ of the training
data. One of the fundamental results in the paper
states that achieving $\epsilon$ learning accuracy (in the PAC model) requires
$\beta \leq \frac{\epsilon}{1+\epsilon}$ for any learning algorithm. For
example, to achieve $90\%$ accuracy ($\epsilon = 0.1$) the adversary
manipulation rate must be less than $10\%$. The efforts below explored this result from a
practical standpoint and introduced poisoning attacks
against ML algorithms. We organize our discussion around the adversarial
capabilities highlighted in the preceding section. Unlike some attacks at inference (see
Section~\ref{ssec:inference-blackbox}), training time attacks require some
degree of knowledge about the learning model, in order to find manipulations of
the data that are damaging to the learned model.

\subsubsec{Label manipulation} When adversaries are only able to modify the
labeling information contained in the training dataset, the attack surface is
limited: they must find the most harmful label given partial or full knowledge
of the learning algorithm ran by the defender. The baseline strategy is to
randomly perturb the labels, i.e. select a new label for a fraction of the
training data by drawing from a random distribution. Biggio et al. showed that
this was sufficient to degrade inference performance of classifiers learned with
SVMs~\cite{biggio2011support}, as long as the adversary randomly flips about
$40\%$ of the training labels. It is unclear whether this attack would
generalize to multi-class classifiers, with $k>2$ output classes (these results
only considered problems with $k=2$ classes, where swapping the labels is
guaranteed to be very harmful to the model). The authors also demonstrate that
perturbing points classified with confidence by the model in priority is a
compelling heuristic to later degrade the model's performance during inference.
It reduces the ratio of poisoned points to $30\%$ for comparable drops in
inference accuracy on the tasks also used to evaluate random swaps. A similar
attack approach has been applied in the context of
healthcare~\cite{mozaffari2015systematic}. As is the case for the approach
in~\cite{biggio2011support}, this attack requires that a new ML model be learned
for each new candidate poisoning point in order to measure the proposed point's
impact on the updated model's performance during inference. This high
computation cost is due to the largely unknown relationship between performance
metrics respectively computed on the training and test data.

\begin{myprop}
	Search algorithms for poisoning points are computationally expensive because of
	the complex and often poorly understood relationship between a model's accuracy
	on training and test distributions. 
\end{myprop}

\subsubsec{Input manipulation} In this threat model, the adversary can
corrupt the input features of training points processed by the learning
algorithm, in addition to its labels. These works assume 
knowledge of the learning algorithm and training set.

\boldpara{Direct poisoning of the learning inputs} Kloft et al. show that by
inserting points in a training dataset used for anomaly detection, they can gradually
shift the decision boundary of a simple centroid model, i.e. a model that
classifies a test input as malicious when it is too far from the empirical mean
of the training data~\cite{kloft2010online}. The detection model is learned in 
an online fashion---new training data is collected at regular intervals and the parameter values $\theta$ are 
computed based on a sliding window of that data.  
Therefore, injection of poisoning data in the training
dataset is a particularly easy task for adversaries in these online settings. 
Poisoning points are found by
solving a linear programming problem that maximizes the displacement of the centroid (empirical mean of the training data). This formulation is made possible by the
simplicity of the centroid model, which essentially evaluates an Euclidean distance.

\begin{myprop}
	The poisoning attack surface of a ML system is often exacerbated
	when learning is performed online, i.e. new training points are added  
	by observing the environment in which the system evolves.
\end{myprop}

In the settings of offline learning, Biggio et al. introduce an attack that also
inserts inputs in the training set~\cite{biggio2012poisoning}. These malicious
samples are crafted using a gradient ascent method that identifies inputs
corresponding to local maxima in the test error of the model. Adding these
inputs to the training set results in a degraded
classification accuracy at inference. Their approach is specific to SVMs,
because it relies on the existence of a closed-form formulation of the model's
test error, which in their case follows from the assumption that support
vectors\footnote{Support vectors are the subset of training points that suffice
	to define the decision boundary of a support vector machine.} do not change as a
result of the insertion of poisoning points. Mei et al. introduce a more general
framework for poisoning, which finds the optimal changes to the training set in
terms of cardinality or the Frobenius norm, as long as the targeted ML model is
trained using a convex loss (e.g., linear and logistic regression or SVMs) and its
input domain is continuous~\cite{mei2015using}. Their attack is formulated as
two nested optimization problems, which are solved by gradient descent after reducing them to a single optimization problem using the inner problem's Karush-Kuhn-Tucker conditions.

\boldpara{Indirect poisoning of the learning inputs} Adversaries
with no access to the pre-processed data must instead poison the
model's training data before its pre-processing (see Figure~\ref{fig:ml-attack-surface}). For instance, Perdisci et al.
prevented Polygraph, a worm signature generation
tool~\cite{newsome2005polygraph}, from learning meaningful signatures by
inserting perturbations in worm traffic
flows~\cite{perdisci2006misleading}. Polygraph combines a flow
tokenizer together with a classifier that determines whether a flow should be
in the signature. Polymorphic worms are crafted with noisy traffic flows
such that (1) their tokenized representations will share tokens not
representative of the worm's traffic flow, and (2) they modify the classifier's
threshold for using a signature to flag worms. This attack forces Polygraph to
generate signatures with tokens that do not correspond to invariants of the
worm's behavior. Later, Xiao et al. adapted the gradient ascent strategy
introduced in~\cite{biggio2012poisoning} to feature selection
algorithms like LASSO~\cite{xiao2015feature}.

\subsection{Targeting Privacy}

During training, the confidentiality and privacy of the data
and ML model are not impacted by the fact that ML is
used, but rather the extent of the adversary's access to the system hosting
the data and model. This is a traditional access control
problem, which falls outside the scope of our discussion.

\section{Inferring in Adversarial Settings}
\label{sec:inference}

Adversaries may also attack ML systems at inference time. In such settings, they cannot poison the training data or
tamper with the model parameters. Hence, the key characteristic that differentiates
attackers is their capability of accessing (but not modifying) the deployed model's internals.
\emph{White-box} adversaries possess knowledge of the internals: e.g., the
ML technique used or the  parameters learned. Instead,
\emph{black-box} access is a weaker assumption corresponding to the capability
of issuing queries to the model or collecting a surrogate training dataset.
Black-box adversaries may surprisingly jeopardize the integrity of the model output, but
white-box access allows for finer-grain control of the outputs. With respect to
privacy, most existing efforts focus on the black-box (oracle) attacks that expose
properties of the training data or the model itself.

\begin{figure*}
	\centering
{\footnotesize
	\begin{tabular}{|c|c|c|c|c|c|c|c|}
		\cline{4-8}
		 \multicolumn{3}{c|}{} & \multicolumn{2}{c|}{Integrity} & \multicolumn{3}{c|}{Privacy} \\ \cline{1-8}
		 Knowledge   & Access to model input  & Access to &  & Source-target &  & Training data & Model   \\ 
		 of model $h_\theta$&  $x$ and output $h(x)$  & training data & Misprediction & misprediction & Membership & extraction & extraction  \\ \hline 
		 \multirow{2}{*}{White-Box} & Full & No & \cite{biggio2013evasion,goodfellow2014explaining,moosavi2015deepfool} & \cite{szegedy2013intriguing,papernot2015limitations,alfeld2016data} &  & \cite{ateniese2015hacking}  &    \\ \cline{2-8}
		  & Through pipeline only & No & \cite{grosse2016adversarial,kurakin2016adversarial,sharif2016accessorize} & \cite{sharif2016accessorize} &  &  &    \\ \hline
		  \multirow{4}{*}{Black-Box} & Yes & No & \cite{xu2016automatically} &  & \cite{shokri2016membership} & \cite{fredrikson2014privacy}  & \cite{tramer2016stealing}   \\ \cline{2-8}
		  & \multirow{2}{*}{Input $x$ only}  &  Yes & \cite{laskov2014practical,szegedy2013intriguing,goodfellow2014explaining} &  &  &   &    \cite{tramer2016stealing}\\ \cline{3-8} 
		&  &  No & \cite{papernot2016practical,papernot2016transferability} &  &  &   &    \\ \cline{2-8} 
		  & Through pipeline only & No & \cite{kurakin2016adversarial} &  &  &   &    \\ \hline
	\end{tabular}
}
	\caption{\textbf{Attacks at inference:} all of these works are discussed in Section~\ref{sec:inference} and represent the threat models explored by the research community.}
	\label{tbl:attacks-inference}
\end{figure*}

\subsection{White-box adversaries} \label{inferring-whitebox}

White-box adversaries have varying degrees of access to the model $h$ as well as its parameters $\theta$. This strong threat model allows the
adversary to conduct particularly devastating attacks. While it is often
difficult to obtain, white-box access is not always unrealistic. For instance,
ML models trained on data centers are compressed and deployed to 
smartphones~\cite{hinton2015distilling}, in which case reverse engineering may
enable adversaries to recover the model's internals and thus obtain white-box access.

\subsubsecb{Integrity} To target a white-box system's prediction integrity,
adversaries perturb the ML model inputs. In the theoretical
PAC model, this can be interpreted as modifying the distribution that
generates data during inference. Our discussion of attacks against classifiers
is two-fold: (1) we describe
strategies that require \emph{direct} manipulation of model inputs, and (2) we
consider indirect perturbations \emph{resilient to the pre-processing stages} of the
system's data pipeline. Although most of the research efforts study classification tasks,
we conclude with a discussion of regression and reinforcement learning.

\boldpara{Direct manipulation of model inputs} Here, adversaries alter the feature values processed by the ML model directly. When the model is a
classifier, the adversary seeks to have it assign a wrong class to
perturbed inputs~\cite{barreno2006can}. Szegedy et al. coined the term
\emph{adversarial example} to refer to such inputs~\cite{szegedy2013intriguing}.
Similar to concurrent work~\cite{biggio2013evasion}, they formalize the search
for adversarial examples as the following minimization problem: \begin{equation}
\label{eq:szegedy} \arg\min_r h(x+r) = l \text{~~~s.t.~~~} x+r \in D
\end{equation} The input $x$, correctly classified by $h$, is perturbed with $r$
such that the resulting adversarial example $x^*=x+r$ remains in the input
domain $D$ but is assigned the target label $l$. This is a \emph{source-target
misclassification} as the target class $l\neq h(x)$ is
chosen~\cite{papernot2015limitations}. For non-convex models $h$ like DNNs, the
authors apply the L-BFGS algorithm~\cite{liu1989limited} to solve
Equation~\ref{eq:szegedy}. Surprisingly, DNNs with state-of-the-art accuracy on
object recognition tasks are misled by small perturbations $r$.

To solve Equation~\ref{eq:szegedy} efficiently, Goodfellow et al. introduced the
\emph{fast gradient sign method}~\cite{goodfellow2014explaining}. A
linearization assumption reduces the computation of an adversarial example $x^*$
to: \begin{equation} x^* = x+\epsilon \cdot sign(\nabla_{\vec{x}}
J_h(\theta,x,y) ) \end{equation} where $J_h$ is the cost function used to train
the model $h$. Despite the approximation made, a model with close to
state-of-the-art performance on MNIST\footnote{The MNIST dataset~\cite{lecun1998mnist} is a widely used corpus of 70,000 handwritten digits used for validating image processing machine learning systems.} misclassifies $89.4\%$ of this method's
adversarial examples. This empirically validates the hypothesis that erroneous model predictions on 
 adversarial examples are likely due to the linear 
extrapolation made by components of ML models (e.g., individual neurons of a DNN) for inputs far from the training 
data. In its canonical form, the technique is designed for
misclassification (in any class differing from the correct class), but it can be
extended to choose the target class.

\begin{myprop}
Adversarial
examples
exist in half-spaces of the model's output
surface because of the overly linear extrapolation that models, including
non-linear ones, make outside of their training data~\cite{goodfellow2014explaining,warde2016adversarial}.
\end{myprop}

Follow-up work reduced the size of a perturbation $r$ according to different
 metrics~\cite{moosavi2015deepfool,huang2015learning}. Papernot et al. introduced a
Jacobian-based adversarial example crafting algorithm that minimizes the number
of features perturbed, i.e. the $L_0$ norm of
$r$~\cite{papernot2015limitations}. On average, only $4\%$ of the features of an
MNIST test set input are perturbed to have it classified in a chosen target
class with $97\%$ success. This proves essential when the ML model has a
discrete input domain for which only a subset of the features can be modified
easily by adversaries. This is the case of malware detectors: in this application, adversarial
examples are malware applications classified as
benign~\cite{grosse2016adversarial}.

Classifiers always output a class,
even if the input is out of the expected distribution. It is therefore not
surprising that randomly sampled inputs can be constrained to be classified in a
class with confidence~\cite{goodfellow2014explaining,nguyen2014deep}. The security consequences
are not so important since humans would not classify these samples in any of the
problem's classes. Unfortunately, training models with a class specific to
rubbish (out of distribution) samples does not mitigate adversarial
examples~\cite{goodfellow2014explaining}.

\boldpara{Indirect manipulation of model inputs} When the adversary cannot directly modify
feature values used as inputs of the ML model, it must find perturbations that
are preserved by the data pipeline that precedes the classifier in the overall
targeted system. Strategies operating in this threat model construct adversarial
examples in the \emph{physical domain} stage of
Figure~\ref{fig:ml-attack-surface}.

Kurakin et al. showed how printouts of adversarial examples produced by the fast
gradient sign algorithm were still misclassified by an object recognition
model~\cite{kurakin2016adversarial}. They fed the model with photographs of the
printouts, thus reproducing the typical pre-processing stage of a computer
vision system's data pipeline. They also found these physical adversarial
examples to be  resilient to pre-processing deformations like contrast
modifications or blurring. Sharif et al. applied the approach introduced
in~\cite{szegedy2013intriguing} to find adversarial examples that are printed on
glasses frames, which once worn by an individual result in its face being
misclassified by a face recognition model~\cite{sharif2016accessorize}. Adding
penalties to ensure the perturbations are physically realizable (i.e.,
printable) in Equation~\ref{eq:szegedy} is sufficient to conduct
misclassification attacks (the face is misclassified in any wrong class), and to
a more limited extent source-target misclassification attacks (the face is
misclassified in a chosen target class).

\begin{myprop}
	To be resilient to the pipeline's deformations, adversarial examples
	in physical domains need to introduce adapted, often larger, perturbations.
\end{myprop}

As a natural extension to~\cite{nguyen2014deep} (see above), it
was shown that rubbish audio can be classified with confidence by a
speech recognition system~\cite{carlini2016hidden}. Consequences are
not as important in terms of security then~\cite{sharif2016accessorize}: the audio does not
correspond to any legitimate input expected by the speech system or humans.

\boldpara{Beyond classification} While most work has focused on attacking
classifiers, Alfeld et al.~\cite{alfeld2016data} look at autoregressive models.
An autoregressive model is one where the prediction $x_t$ of a time series
depends on previous $k$ realizations of $x$, that is, $x_t = \sum_{i=1}^k c_i
x_{t - i}$; such models are widely used in market predictions. The adversary can
manipulate the input data with the goal of achieving their desired prediction,
given budget constraints for the adversary. The author's formulate the
adversary's manipulation problem as a quadratic optimization problem and provide
efficient solutions for it. 

Adversarial behaviors were also considered in reinforcement learning, albeit not
to address security but rather to improve the utility of models. Pinto
et al. improve a model for grasping objects by introducing a competing model
that attempts to snatch objects before the original model successfully grasps
them~\cite{pinto2016supervision}. The two models are trained, \`a la Generative Adversarial Networks~\cite{goodfellow2014generative}, with competitive cost functions.

\begin{myprop}
	Although research has focused on classification problems, algorithms 
	developed to craft adversarial examples naturally extend to other settings like
	reinforcement learning: e.g., the adversary perturbs a video game frame to force
	an agent to take wrong actions.
\end{myprop}

\subsubsecb{Privacy} As discussed in Section~\ref{sec:ml-security}, adversaries targeting the 
privacy of a ML system are commonly interested in  
recovering information about either the training data or the learned model itself.
The simplest attack consists in 
performing a membership test, i.e. determining whether a particular input was
used in the training dataset of a model. Stronger opponents may seek to
extract fully or partially unknown training points. Few attacks operate
in the white-box threat model, as the black-box model (see below) is more realistic
when considering privacy. 

Ateniese et al. infer statistical information 
about the training dataset from a trained model $h_\theta$~\cite{ateniese2015hacking}: i.e., they analyze the model to
determine whether its training data verified a certain statistical property. 
Their attack 
generates several datasets, where some exhibit
the statistical property and others don't. A model is trained on each dataset independently. The adversary then trains a \emph{meta-classifier}:
it takes as its inputs these models and predicts if their dataset verified the statistical property. The meta-classifier is finally applied to the model of interest $h_\theta$ to fulfill the initial adversarial goal. One limitation is 
that all classifiers must be trained with the same technique than $h_\theta$.

\subsection{Black-box adversaries}
\label{ssec:inference-blackbox}

When performing attacks against \emph{black-box} systems, adversaries do not
know the model internals. This prohibits the strategies described in
Section~\ref{inferring-whitebox}: for instance, integrity attacks require that
the attacker compute  gradients defined using the model $h$ and its parameters
$\theta$. However, black-box access is perhaps a more realistic threat model, as all it requires is access to the output responses. For instance, an adversary
seeking to penetrate a computer network rarely has access to the specifications
of the intrusion detection system deployed--but they can often observe how it  responds to network events. 
Similar attacks are key to performing reconnaissance in networks to determine their environmental detection and response policies. We focus on strategies designed
irrespectively of the domain ML is being applied to, albeit
heuristics specific to certain applications exist~\cite{wittel2004attacking,lowd2005good}, e.g., spam filtering.

A common threat model for black-box adversaries is the one of an \emph{oracle}, borrowed from the
cryptography community: the adversary may issue queries to the ML model and
observe its output for any chosen input. This is particularly relevant in
the increasingly popular environment of ML as a Service cloud platforms, where the model is potentially accessible
through a query interface. A PAC based work shows that with no
access to the training data or ML algorithm, querying the target
model and knowledge of the class of target models allows the adversary to
reconstruct the model with similar amount of query data as used in
training~\cite{vorobeychik2014optimal} Thus, a key metric when comparing
different attacks is the wealth of information returned by the oracle, and
the number of oracle queries.

\subsubsec{Integrity} Lowd et al. estimate the cost of misclassification in
terms of the number of queries to the black-box model~\cite{lowd2005adversarial}. The
adversary has oracle access to the model. A cost function is associated with
modifying an input $x$ to a target instance $x^*$. The cost function is a
weighted $l_1$difference between $x^*$ and $x$.
The authors
introduce ACRE learnability, which poses the problem of finding the least cost
modification to have a malicious input classified as benign using a
polynomial number of queries to the ML oracle. It is shown that continous
features allow for ACRE learnability while discrete features make the problem
NP-hard. Because ACRE learnability also depends on the
cost function, it is a different problem from reverse engineering the model.
Following up on this thread, Nelson et al.~\cite{nelson2012query} identify the
space of convex inducing classifiers---those where one of the classes is a
convex set---that are ACRE learnable but not necessarily reverse engineerable.

\boldpara{Direct manipulation of model inputs} It has been hypothesized that in classification, adversaries with 
access to class probabilities for label outputs are only slightly weaker than white-box adversaries. In these settings, Xu et al. apply a computationally expensive genetic algorithm. The
fitness of genetic variants obtained by mutation is defined in terms of the
oracle's class probability predictions~\cite{xu2016automatically}. The approach
evades a random forest and SVM used for malware detection.

When the adversary cannot access probabilities, it is more difficult to extract
information about the decision boundary, a pre-requisite to find input
perturbations that result in erroneous predictions. In the
following works, the adversary only observes the first and last stage of the pipeline
from Figure~\ref{fig:ml-attack-surface}: e.g., the input (which they produce) and the class label in
classification tasks. Szegedy et al. first observed \emph{adversarial example
transferability}: i.e., the property that adversarial examples crafted to be
misclassified by a model are likely to be misclassified by a different
model. This transferability property holds even when models are
trained on different datasets.

Assuming the availability of surrogate data to the adversary, Srndic et al.
explored the strategy of training a substitute model for the targeted
one~\cite{laskov2014practical}. They exploit a semantic gap to evade a malware PDF detector: they inject
additional features that are not interpreted by PDF renderers. As such, their
attack does not generalize well to other application domains or models.

Papernot et al. used the cross-model transferability of adversarial
samples~\cite{szegedy2013intriguing,goodfellow2014explaining} to design a
black-box attack~\cite{papernot2016practical}. They demonstrated
how attackers can force a remotely hosted ML model to misclassify inputs without
access to its architecture, parameters, or training data. The attack trains a substitute model using synthetic inputs generated by
the adversary and labeled by querying the oracle. The substitute model is then
used to craft adversarial examples that transfer back to---are misclassified
by---the originally targeted model. They force a DNN trained by MetaMind, an
online API for deep learning, to misclassify inputs at a rate of $84.24\%$. In a
follow-up work~\cite{papernot2016transferability}, they show that the
attack generalizes to many ML models by having a logistic regression oracle
trained by Amazon misclassify $96\%$ of the adversarial examples crafted. 

\begin{myprop}
	Black-box attacks make it more difficult for the adversary to choose a target class in which the perturbed input will be classified by the model, when compared to white-box settings.
\end{myprop}

\boldpara{Data pipeline manipulation} Using transferability, Kurakin et al.~\cite{kurakin2016adversarial} demonstrated that physical adversarial example (i.e., printouts of an adversarial example) can also mislead an object recognition model included in a smarphone app, which differs from the one used to craft the adversarial example.
These findings suggest that black-box adversaries are able to craft
inputs misclassified by the ML model despite the pre-processing stages of the
system's data pipeline.

\subsubsec{Privacy} In black-box settings, adversaries targeting privacy may
pursue the goals already discussed in white-box settings: membership attacks
and training data extraction. In addition, since the model internals are now unknown to them, extracting model parameters themselves
is now a valid goal.
 
\boldpara{Membership attacks} This type of adversary is looking to test whether
or not a specific point was part of the training dataset analyzed to learn the
model's parameter values. Shokri et al. show how to conduct this type of attack,
named \emph{membership inference}, against black-box
models~\cite{shokri2016membership}. Their strategy exploits differences in the
model's response to points that were or were not seen during training. For each
class of the targeted black-box model, they train a shadow model, with the same
ML technique. Each shadow model is trained to solve the membership inference
test for samples of the corresponding class. The procedure that generates
synthetic data is initialized  with a random input and
performs hill climbing by querying the original model to find modifications of
the input that yield a classification with strong confidence in a class of the
problem. These synthetic inputs are assumed to be statistically similar to the
inputs contained in the black-box model's training dataset.

\boldpara{Training data extraction} 
Fredrikson et al. present the model inversion
attack~\cite{fredrikson2014privacy}. For a medicine dosage prediction task, they
show that given access to the model and auxiliary information about the
patient's stable medicine dosage, they can recover genomic information about the
patient. Although the approach illustrates privacy concerns that may arise from
giving access to ML models trained on sensitive data, it is unclear whether the
genomic information is recovered because of the ML model or the strong
correlation between the auxiliary information that the adversary also has access
to (the patient's dosage)~\cite{mcsherry2016statistical}.  Model inversion
enables adversaries to extract training data from observed model
predictions~\cite{fredrikson2015model}. However, the input extracted is not
actually a specific point of the training dataset, but rather an average
representation of the inputs that are classified in a class---similar to what is
done by saliency maps~\cite{zeiler2014visualizing}. The demonstration is
convincing in~\cite{fredrikson2015model} because each class corresponds to a
single individual.

\boldpara{Model extraction} Among other considerations like intellectual property, extracting ML model has privacy implications---as models have been shown to memorize training data at least partially. Tramer et al. show how to extract parameters of a model from the observation of its predictions~\cite{tramer2016stealing}. Their attack is conceptually simple: it consists in applying equation solving to recover parameters $\theta$ from sets of observed input-output pairs $(x,h_\theta(x))$. However, the approach does not scale to scenarios where the adversary looses access to the probabilities returned for each class, i.e. when it can only access the label. This leaves room for future work to improve upon such extraction techniques to make them practical.

\section{Towards Robust, Private, and Accountable Machine Learning Models}
\label{sec:future}

After presenting attacks conducted at training
in Section~\ref{sec:training} and inference in
Section~\ref{sec:inference}, we cover efforts at the
intersection of security, privacy, and ML that are
relevant to the mitigation of these previously discussed attacks. 
We draw parallels between the seemingly
unrelated goals of: (a) robustness to distribution drifts, (b) learning
privacy-preserving models, and (c) fairness and accountability. 
Many of these remain largely open problems, thus we draw insights useful for future work.

\subsection{Robustness of models to distribution drifts}
\label{ssec:future-robust}

To mitigate the integrity attacks presented in Section~\ref{sec:inference}, ML
needs to be robust to \emph{distribution drifts}: i.e., situations where the
training and test distributions differ. Indeed, adversarial manipulations are
instances of such drifts. During inference, an adversary might introduce
positively connotated words in spam emails to evade detection, thus creating a
test distribution different from the one analyzed during
training~\cite{lowd2005good}. The opposite, modifying the training distribution,
is also possible: the adversary might include an identical keyword in many spam
emails used for training, and then submit spam ommiting that keyword at test
time. Within the PAC framework, a distribution drift violates the assumption
that more training data reduces the learning algorithm's error rate. We include
a PAC-based analysis of learning robust to distribution drifts.

\subsubsec{Defending against training-time attacks} 
Most defense mechanism at training-time rely on the fact that poisoning samples are typically out of the expected input distribution. 

Rubinstein et al.~\cite{rubinstein2009antidote} pull from robust statistics to
build a PCA-based detection model robust to poisoning.
To limit the influence of outliers to the training distribution, they
constrain the PCA algorithm to search for a direction whose projections maximize
a univariate dispersion measure based on robust projection pursuit estimators
instead of the standard deviation. In a similar approach, Biggio et al. limit the
vulnerability of SVMs to training label manipulations by adding a
regularization term to the loss function, which in turn reduces the model
sensitivity to out-of-diagonal kernel matrix elements~\cite{biggio2011support}.
Their approach does not impact the convexity of the optimization problem 
unlike previous attempts~\cite{stempfel2009learning, xu2006robust}, which reduces the
impact of the defense mechanism on performance.

Barreno et al. make proposals to secure learning~\cite{barreno2006can}.
These include the use of regularization in the optimization
problems solved to train ML models. This removes some of the complexity
exploitable by an adversary (see Section~\ref{sec:learning-theory}). 
The authors also mention an attack detection strategy based
on isolating a special holdout set to detect poisoning attempts.
Lastly, they suggest the use of disinformation
with for instance honeypots~\cite{provos2004virtual} and randomization of the
ML model behavior.

\subsubsec{Defending against inference-time attacks} The difficulty in attaining robustness to adversarial manipulations at
inference, i.e. malicious test distribution drifts, stems from
the inherent complexity of output surfaces learned by ML techniques. Yet, a
paradox arises from the observation that this complexity of ML hypotheses is
necessary to confer modeling capacity sufficient to train robust models (see Section~\ref{sec:learning-theory}).
Defending against attacks at inference remains
largely an open problem. We 
explain why mechanisms that smooth model outputs in infinitesimal
neighborhoods of the training data fail to guarantee integrity. Then, we
present more effective strategies that make models robust to larger perturbations of their inputs

\boldpara{Defending by gradient masking} Most integrity attacks 
in Section~\ref{sec:inference} rely on the adversary being able to find small
perturbations that lead to significant changes in the model output. Thus, a
natural class of defenses seeks to reduce the sensitivity of models to
small changes made to their inputs. This sensitivity is estimated by computing
first order derivatives of the model $h$ with respect to its inputs. These
gradients are minimized during the learning phase: hence the \emph{gradient
	masking} terminology. We detail why this intuitive strategy is bound
to have limited success because of adversarial example transferability.

Gu et al. introduce a new ML model, which they name \emph{deep contractive
	networks}, trained using a smoothness penalty~\cite{gu2014towards}. The penalty
is defined with the Frobenius norm of the model's Jacobian matrix, and is
approximated layer by layer to preserve computational efficiency. This approach
was later generalized to other gradient-based penalties
in~\cite{lyu2015unified,ororbia2016unifying}. Although Gu et al. show that
contractive models are more robust to adversaries, the 
penalty greatly reduces the capacity of these models, with consequences
on their performance and applicability.

The approach introduced in~\cite{papernot2015distillation} does not
involve the expensive computation of gradient-based penalties. The technique is an adaptation of
\emph{distillation}~\cite{hinton2015distilling}, a mechanism designed
to compress large models into smaller ones while preserving
prediction accuracy. In a nutshell, the large model labels data with class
probabilities, which are then used to train the small model. Instead of compression,
the authors apply distillation to increase the robustness of DNNs to adversarial
samples. They report that the additional entropy in
probability vectors (compared to labels) yields models with smoother output surfaces. In experiments
with the fast gradient sign method~\cite{papernot2016effectiveness} and the
Jacobian attack~\cite{papernot2015distillation}, larger
perturbations are required to achieve misclassification of
adversarial examples by the distilled model. However,
\cite{carlini2016defensive} identified a variant of the attack of~\cite{papernot2015limitations} which distillation
fails to mitigate on one dataset.

A simpler variant of distillation, called label
smoothing~\cite{szegedy2015rethinking}, improves robustness to
adversarial samples crafted using the fast gradient sign
method~\cite{WardeFarley16}. It replaces hard class
labels (a vector where the only non-null element is
the correct class index) with soft labels (each class is assigned a
value close to $1/N$ for a $N$-class problem). However this
variant was found to not defend against more computation expensive but precise
attacks like the Jacobian-based iterative attack~\cite{papernot2015limitations}.

\begin{figure}
	\centering
	\includegraphics[width=\columnwidth]{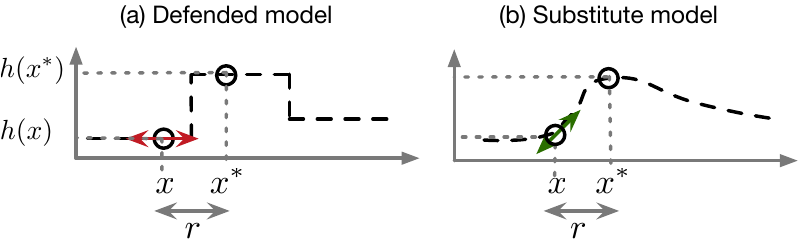}
	\vspace*{-0.3in}
	\caption{\textbf{Evading infinitesimal defenses using transferability:} the
		defended model is very smooth in neighborhoods of training points: i.e., gradients of the model outputs with respect to its inputs are zero
		and the adversary does not know in which direction to look for adversarial
		examples. However, the adversary can use the substitute model's gradients to
		find adversarial examples that transfer back to the defended model. Note that
		this effect would be exacerbated by models with more than one dimension.}
	\label{fig:evading-infinitesimal}
\end{figure}

These results suggest limitations of defense strategies that seek to conceal 
gradient-based information exploited by adversaries. In fact, Papernot et al. report that
defensive distillation can be evaded using a black-box
attack~\cite{papernot2016practical}. We here detail the reason behind this
evasion. When applying defense mechanisms that smooth a model's output surface,
as illustrated in Figure~\ref{fig:evading-infinitesimal}.(a), the adversary
cannot craft adversarial examples because the gradients it needs to compute
(e.g., the derivative of the model output with respect to its input) have values
close to zero. In~\cite{papernot2016practical}, this is referred to as
\emph{gradient masking}. The adversary may instead use a substitute model, illustrated in Figure~\ref{fig:evading-infinitesimal}.(b), to
craft adversarial examples, since the substitute is not impacted by the
defensive mechanism and will still have the gradients necessary to find
adversarial directions. Due to the adversarial example transferability
property~\cite{szegedy2013intriguing} described in Section~\ref{sec:inference},
the adversarial examples crafted using the substitute are also misclassified by
the defended model. This attack vector is likely to apply to any defense
performing gradient masking, i.e. any mechanism defending against adversarial
examples in infinitesimal neighborhoods of the training points.

\begin{myprop}
	Any defense that tampers with adversarial example crafting heuristics (e.g., by
	masking gradients used by adversaries) but does not mitigate the underlying
	erroneous model predictions can be evaded using a transferability-based
	black-box attack.
\end{myprop}

\boldpara{Defending against larger perturbations} Szegedy et
al.~\cite{szegedy2013intriguing} first suggested injecting
adversarial samples, correctly labeled, in the training set as a means to make the model robust. They showed that models fitted with
this mixture of legitimate and adversarial samples were regularized and more
robust to future adversaries. 
The efficiency of the fast gradient sign
method allows for the integration of an adversarial objective during training. The adversarial objective minimizes the error
between the model's prediction on the adversarial example  and the original sample label. This \emph{adversarial training}
continuously makes the model more robust to adversarial examples crafted with the latest model parameters. Goodfellow et al. show that this
reduces the misclassification rate of a MNIST
model from $89.4\%$ to $17.9\%$ on adversarial examples~\cite{goodfellow2014explaining}.

Huang et al.~\cite{huang2015learning} developed the intuition behind adversarial training, i.e. penalize misclassification of adversarial examples.
They formulate a min-max problem between the adversary
applying perturbations to each training point to maximize the model's
classification error, and the learning procedure attempting to minimize the
model's misclassification error:
\begin{equation}
\min_h \sum_i \max_{\|r^{(i)}\| \leq c} l(h(x_i+ r^{(i)}),y_i)
\end{equation}
They solve the problem using stochastic gradient
descent~\cite{bottou2010large}. Their experimentation shows 
improvements over~\cite{goodfellow2014explaining}, but they are often statistically
non-significative. The non-adaptiveness of adversarial training explains some of the results reported by Moosavi et al.~\cite{moosavi2015deepfool}, where they apply the defense with an attack and evaluate robustness with another one.

\begin{myprop}
In adversarial training, it is essential to include adversarial
examples produced by all known attacks, as the defensive training is non-adaptive. 
\end{myprop}

\subsubsec{Interpreting robust learning in the PAC model}
As stated earlier, inference attacks can be interpreted in the PAC model  as the adversary choosing a different data distribution during inference from the one used in training.
Thus, an approach to handle such adversaries is to modify the distribution $D$ that is used to generate the training data so that training data samples reflect a probability distribution that maybe encountered during inference, i.e., a distribution that places more probability mass on possibly adversarially classified data. The additions of adversarial samples in the training data~\cite{szegedy2013intriguing,moosavi2015deepfool} or modifying the training loss function~\cite{huang2015learning} can be viewed as modifying the training distribution $D$ towards such an end.

Recall that the PAC model captures the fact that learning algorithms only optimize for expected loss and hence existence of mis-classfied instances can never be ruled out completely. Thus, a formal approach of modifying the training data must also consider the adversary's cost in modifying the distribution in order to tractably deal with adversarial manipulation. Game theory is a natural tool to capture such defender-adversary interaction. Next, we use the Stackelberg game framework to capture the adversarial interaction. It is a model for defender-adversary interaction where the defender lays out her strategy (in this paper the classifier) and the adversary responds optimally (choose a least cost evasion). Stackelberg games also allow for scalability compared to the corresponding simultaneous move game~\cite{sinha2015physical,tambe2011security}.

Bruckner et al.~\cite{bruckner2011stackelberg} first recognized that test data manipulation can be seen in the PAC framework as modifying the distribution $D$ to $\overset{\boldsymbol .}{D}$. Then, the learner's expected cost for output $h$ is $E_{x,y \sim \overset{\boldsymbol .}{D}}[l_h(x,y)]$. 

The basic approach in all game based adversarial learning technique is to associate a cost function $c: X \times X \rightarrow \mathbb{R}$ that provides the cost $c(x, x')$ for the adversary in modifying the feature vector $x$ to $x'$. The game involves the following two stages (we provide a generalization of the statistical classification game by Hardt et al.~\cite{Hardt2016}):
\begin{enumerate}
\item The defender publishes its hypothesis $\hat{h}$. She has knowledge of $c$, and training samples drawn according to $D$.
\item The adversary publishes  a modification function $\Delta$.
\end{enumerate}
The defender's loss is $E_{x,y \sim D}[l_h(\Delta(x),y)]$ and the adversary's cost is $E_{x,y \sim D}[l_h^a(\Delta(x),y) + c(x, \Delta(x))]$ where $l_h^a(x)$ is a loss function for the adversary that captures the loss when the prediction is $h(x)$ and the true output is $y$. It is worth pointing out that $\Delta: X \rightarrow X$ can depend on the hypothesis $\hat{h}$. This game follows the test data manipulation framework described earlier. The function $\Delta$ induces a change of the test distribution $D$ to some other distribution $\overset{\boldsymbol .}{D}$\footnote{If $(X, Y)$ is a random value distributed according to $D$, then the distribution of $(\Delta(X), Y)$ is $\overset{\boldsymbol .}{D}$.} and the defender's loss function can be written as $E_{x,y \sim \overset{\boldsymbol .}{D}}[l_h(x,y)]$. Thus, just like the PAC framework, the costs for both players are stated in terms of the unknown $D$ (or $\overset{\boldsymbol .}{D}$) and then the empirical counterparts of these functions are:
$$
\frac{1}{n}\sum_{i=1}^n l_h(\Delta(x_i),y_i) \mbox{ and }
\frac{1}{n}\sum_{i=1}^n l_h^a(\Delta(x_i),y_i) + c(x_i, \Delta(x_i)) 
$$
The Stackelberg equilibrium computation problem is stated below. This problem is the analogue of the empirical risk minimization that the learner solves in the standard setting.
\begin{align*}
\min & \sum_{i=1}^n l_h(\Delta(x_i),y_i) \\
& \Delta \in \arg\!\min_{\Delta} \sum_{i=1}^n l_h^a(\Delta(x_i),y_i) + c(x_i, \Delta(x_i)) 
\end{align*}
Unlike the standard empirical risk minimization, this problem is a bi-level optimization problem and is in general NP Hard. Bruckner et al.~\cite{bruckner2011stackelberg} add a regularizer for the learner and the cost function $c(x, x')$ as the $l_2$ distance between $x$ and $x'$. They solve the problem with a sequential quadratic approach.

Following the approach of Lowd et al.~\cite{lowd2005adversarial}, Li et al.~\cite{li2014feature} use a cost function that is relevant for many security settings. The adversary is interested in classifying a malicious data point as non-malicious. Thus, the cost function only imposes costs for modifying the malicious data points.

\subsection{Learning and Inferring with Privacy}

One way of defining privacy-preserving models is that they do not reveal any
additional information about the subjects involved in their training data. This
 is captured by \emph{differential
	privacy}~\cite{dwork2014algorithmic}, a rigorous framework to analyze the
privacy guarantees provided by algorithms. Informally, it formulates privacy as
the property that an algorithm's output does not differ significantly
statistically for two versions of the data differing by only one record. In our
case, the record is a training point and the algorithm the ML model.

A randomized algorithm is said to be $(\varepsilon, \delta)$ differentially
private if for two neighboring training datasets $T, T'$, i.e. which differ by
at most one training point, the algorithm $A$ satisfies for any acceptable
set $S$ of  algorithm outputs: \begin{equation} \label{eq:dp} Pr[A(T)\in S] \leq
e^\varepsilon Pr[A(T')\in S] + \delta \end{equation} The
parameters $(\varepsilon, \delta)$ define an upper bound on the probability that
the output of $A$ differs between $T$ and $T'$. Parameter
$\varepsilon$ is a privacy budget: smaller budgets yield
stronger privacy guarantees. The second parameter $\delta$ is a failure rate
for which it is tolerated that the bound defined by $\varepsilon$ does
not hold.

\subsubsec{Training} The behavior of a ML system needs to be randomized in
order to provide privacy guarantees.
At
training, random noise may be injected to the data, the cost minimized by the
learning algorithm, or the values of parameters learned. 

An instance of training data randomization 
is formalized by local privacy~\cite{kasiviswanathan2011can}. 
In the scenario where users send reports to 
a centralized server that trains a model with the data collected, \emph{randomized response} protects privacy: users respond to server queries with the true answer at a probability $q$, and otherwise return a random value with probability $1-q$. Erlingsson et al. showed that this allowed the developers of a browser to collect meaningful and privacy-preserving usage statistics from users~\cite{erlingsson2014rappor}. Another way to obtain
randomized training data is to first learn an ensemble of teacher models on data partitions, and then use these models to make noisy predictions on public unlabeled data, which is
used to train a private student model. This strategy was explored in~\cite{hamm2016learning,papernot2016semi}.

Chaudhuri et al. show that \emph{objective perturbation}, i.e. introducing random noise---drawn from an exponential distribution and scaled using the model sensitivity\footnote{In differential privacy research,
		sensitivity denotes the maximum change in the model output when one
		training point is changed. This is not identical to the
		sensitivity of ML models to adversarial perturbations (see
		Section~\ref{sec:inference}).}---in the  cost function minimized during learning, can provide $\varepsilon$-differential
privacy~\cite{chaudhuri2011differentially}. Bassily et al. provide improved algorithms and privacy analysis, along with references to many of the works intervening in private empirical risk minimization~\cite{bassily2014differentially}

\begin{myprop}
	Learning models with differential privacy guarantees is difficult because the sensitivity of models is unknown for most interesting ML techniques.
\end{myprop}

Despite noise injected in parameters, Shokri et al. showed 
that large-capacity models like deep neural networks can be trained with  multi-party computations
from perturbed parameters and provide differential privacy guarantees ~\cite{shokri2015privacy}. 
Later, Abadi et al. introduced an alternative approach to improve the privacy
guarantees provided: the strategy followed is to randomly perturb
parameters during the stochastic gradient descent performed by the learning algorithm~\cite{abadi2016deep}. 

\subsubsec{Inference} To provide differential privacy, the ML's behavior may also be randomized
at inference by introducing noise to predictions. Yet, this degrades the accuracy of predictions, since the amount of noise introduced increases with the number of inference queries answered by the ML model. Note that different forms of privacy can be provided during inference. For instance, Dowlin et al. use homomorphic encryption~\cite{rivest1978data} to encrypt the data in a form that allows a neural network to process it without decrypting it~\cite{gilad2016cryptonets}. Although, this does not provide differential privacy, it protects the confidentiality of each individual input. The main limitations are the performance overhead and the restricted set of arithmetic operations supported by homomorphic encryption, which introduce additional constraints in the architecture design of the ML model.

\subsection{Fairness and Accountability in Machine Learning}

The opaque nature of ML generates concerns  regarding a lack of fairness and accountability of decisions taken based on the model
predictions. This is especially important in applications like credit decisions
or healthcare~\cite{international2009estimation}.

\subsubsec{Fairness} In the pipeline from Figure~\ref{fig:ml-attack-surface},
\emph{fairness} is relevant to the action taken in the physical
domain based on the model prediction. It should not nurture discrimination
against specific individuals~\cite{kleinberg2016inherent}. Training
data is perhaps the strongest source of bias in ML. For instance, a dishonest
data collector might adversarially attempt to manipulate the learning into
producing a model that discriminates against certain groups. Historical data 
also inherently reflects social biases~\cite{barocas2016big}. To learn fair
models, Zemel et al. first learn an intermediate
representation that encodes a sanitized variant of the
data, as first discussed in~\cite{dwork2012fairness}. Edwards et al. showed that fairness could be
achieved by learning in competition with an adversary trying to predict the
sensitive variable from the fair model's prediction~\cite{edwards2015censoring}.
They find connections between fairness and privacy, as their approach also
applies to the task of removing sensitive annotations from images. We
expect future work at the intersection of fairness and topics discussed in
this paper to be fruitful.

\subsubsec{Accountability} \emph{Accountability}
explains model predictions using the ML model
internals $h_\theta$. This is fundamentally relevant to
understanding model failures on adversarial examples. Few  models are
interpretable by design, i.e., match human
reasoning~\cite{letham2015interpretable}. Datta et al. introduced
\emph{quantitative input influence} measures to estimate the influence of
specific inputs on the model output~\cite{datta2016algorithmic}. Another avenue
to provide accountability is to compute inputs  that the machine learning
model's components are most sensitive to. An
approach named \emph{activation maximization} synthesizes an input that highly
activates a specific neuron in a neural network~\cite{erhan2009visualizing}. 
The challenge lies in producing synthetic inputs easily interpreted by humans~\cite{mahendran2016visualizing} but faithfully representing the model's behavior.

\section{No Free Lunch in Adversarial Learning}
\label{sec:learning-theory}

\begin{figure}[t]
  \centering
    \includegraphics[width=0.75\linewidth]{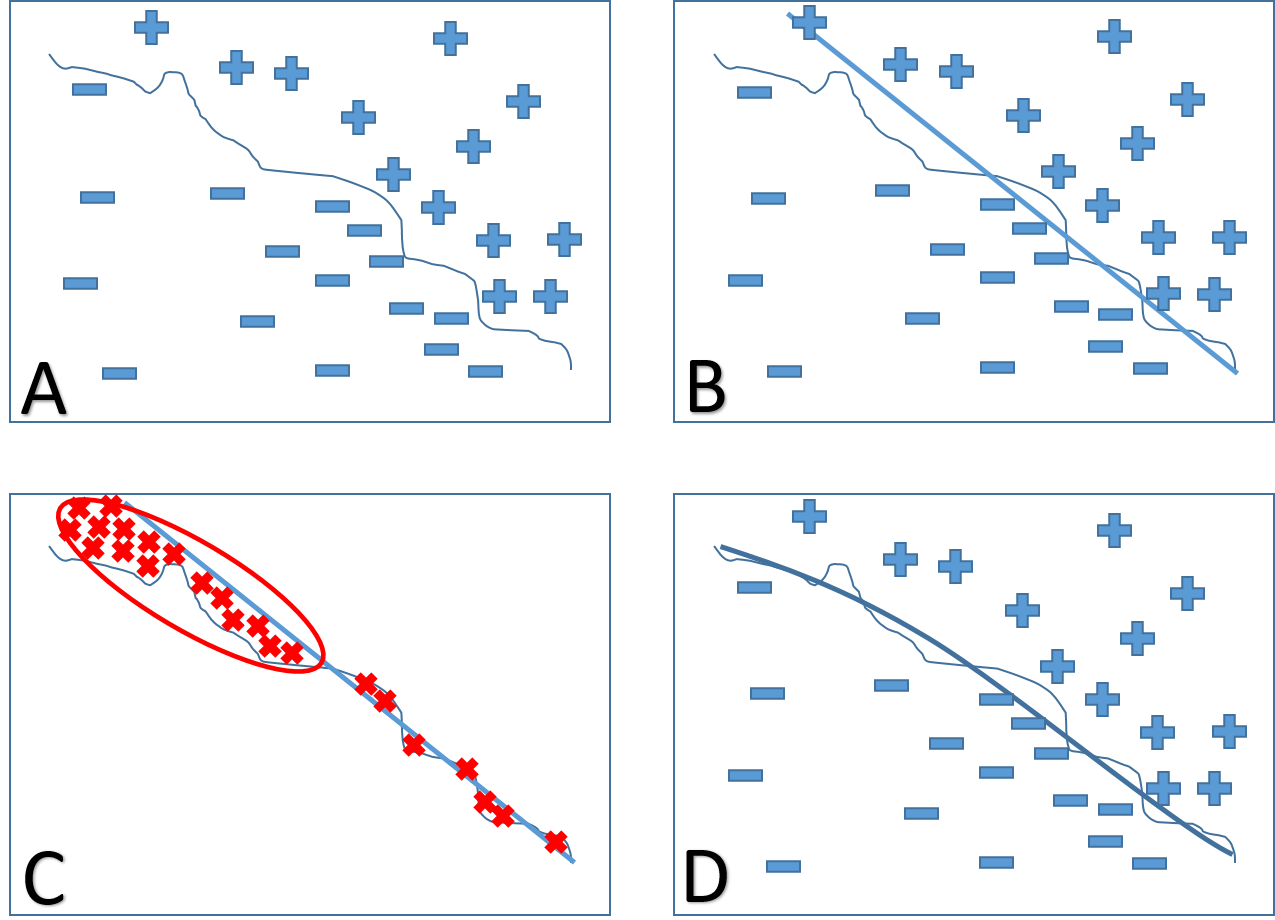}\label{NFLfig}
    \vspace*{-0in}
    \caption{Subfigure A shows the data available for learning and the true separator between the positive and negative region. The top left corner has few data points, which in the PAC model means that the data distribution $D$ has low probability mass over that region. Subfigure B shows the model learned with an hypothesis class $\mathcal{H}$ of linear classifiers. Subfigure C shows all points misclassified by the linear model. Also shown is an adversarially chosen uniform distribution $\overset{\boldsymbol .}{D}$ restricted to the red oval in the top left corner; two observations are (1) the red crosses will cause a significant prediction loss with $\overset{\boldsymbol .}{D}$ and the linear model shown, and (2) the true separator in this red oval is highly non-linear (compared to rest of the space) and hence even the best linear classifier learned w.r.t. $\overset{\boldsymbol .}{D}$ will suffer significant expected loss. Subfigure D shows that a more complex non-linear classifier can be more accurate and can provide a lower expected loss against $\overset{\boldsymbol .}{D}$ (modulo over-fitting issues).}
\end{figure}

We begin by pointing out that if a classifier is perfect, i.e., predicts  the right class for every possible input, then it cannot be manipulated. Thus, the presence of adversarial examples is a manifestation of the classifier being inaccurate on many inputs. Hence the dichotomy between robustness to adversarial examples and better prediction is a false one. Also, it is well-known in ML that, given enough data, more complex hypothesis classes (e.g., non-linear classifier as opposed to linear ones) provide better prediction (see Figure~\ref{NFLfig}). As a result, we explore the interaction between prediction loss, adversarial manipulation at inference and complexity of hypothesis class.

Recall the theoretical characterization for data poisoning by Kearns and Li~\cite{kearns1993learning} (see Section~\ref{sec:training}). While poisoning attacks can be measured by the percentage of data modified, mathematically describing an attack at inference is non-obvious. Thus, our \emph{first result} in this section is an identification of the characteristics of an \emph{effective attack} at inference. Our \emph{second result} reveals that, given a positive probability of presence of an adversary, any supervised ML algorithm suffers from performance degradation under an effective attack. Finally, our \emph{third result} is 
that increased capacity is required for resilience to adversarial examples (and can also give more precision as a by-product). But to prevent empirical challenges, e.g., overfitting, more data is needed to accompany the increase in capacity. Yet, in most practical settings, one is given a dataset of fixed size, which creates a tension between resilience and precision. A trade-off between the two must be found by empirically searching for the optimal capacity to model the underlying distribution. Note that this result (presented below) is analogous to the no free lunch result in data privacy that captures the tension between utility and privacy~\cite{dwork2008differential,kifer2011no}.

In the PAC learning model, data $x,y$ is sampled from a distribution $D$. Recall that the learner learns $\hat{h}$ such that $
P(|r(h^*) - r(\hat{h})| \leq \epsilon) \geq 1 - \delta
$. For this section, we assume that there is enough data\footnote{This assumption is practical in many applications today. Moreover, insufficient data presents fundamental information theoretic limitations~\cite{anthony2009neural} on learning accuracy problems in the benign (without adversaries) setting itself.} so that $\epsilon, \delta$ are negligible, hence we assume $\hat{h}$ is same as $h^*$ for all practical purpose. Recall that the \emph{performance} of any learning algorithm is characterized by the expected loss of its output $h$: $E_{x,y \sim D}[l_{h}(x,y)]$. 
In the real world, it is often not known whether an adversary will be present or not. Thus, we assume that an adversary is present with probability $q \in (0,1)$; where $q$ is not extremely close to $0$ or $1$ so that both $q$ and $1-q$ are not negligible. Also, recall that an attack in the inference phase is captured by an adversarially modified distribution $\overset{\boldsymbol .}{D}$ of test data.

Our first result is the following definition that characterizes an effective attack by the adversary. 

\begin{itemize}
\item \textbf{$\alpha$-effective attack against $\mathcal{H}$ and $D$}: the best hypothesis $\overset{\boldsymbol .}{h}^*$ in the adversarial setting, i.e., $\overset{\boldsymbol .}{h}^* \in \arg\!\min_{h\in \mathcal{H}} E_{x,y \sim \overset{\boldsymbol .}{D}}[l_h(x,y)]$ still suffers a loss $E_{x,y \sim \overset{\boldsymbol .}{D}}[l_{\overset{\boldsymbol .}{h}^*}(x,y)]$ such that for the best hypothesis in the benign setting ${h}^* \in \arg\!\min_{h\in \mathcal{H}} E_{x,y \sim {D}}[l_h(x,y)]$ $$E_{x,y \sim \overset{\boldsymbol .}{D}}[l_{\overset{\boldsymbol .}{h}^*}(x,y)] = E_{x,y \sim D}[l_{h^*}(x,y)] + \alpha$$ 
\end{itemize}
This definition implies that for $\alpha > 0$ it is not trivial to defend against the modified distribution that the adversary presents compared to the benign scenario, since even the best hypothesis $\overset{\boldsymbol .}{h}^*$ against the modified distribution $\overset{\boldsymbol .}{D}$ suffers a greater loss than the best hypothesis ${h}^*$ in the benign case.

Note that the definition above does not restrict the adversary to choose any particular $\overset{\boldsymbol .}{D}$, that is, the adversary is not restricted to a particular attack. The definition is parametrized by $\mathcal{H}$ and $D$, that is, the attack is effective against the given choice of hypothesis class $\mathcal{H}$ and data distribution $D$. We leave open the research question about whether a mathematical characterization of when such attacks exists or is feasible given the attackers cost--however, we illustrate in Figure~\ref{NFLfig}-C why it is reasonable to assume that such attacks abound in practice.

For the sake of comparison, we contrast our definition with a possible alternate attack definition: one may call an attack effective if $E_{x,y \sim \overset{\boldsymbol .}{D}}[l_{{h}^*}(x,y)] = E_{x,y \sim D}[l_{h^*}(x,y)] + \alpha$, that is, the adversarial choice of $\overset{\boldsymbol .}{D}$ cause an increase in prediction loss against the deployed $h^*$. While such an attack is indeed practical as shown in Figure~\ref{NFLfig}-C, the impact of such an attack can be reduced if a different data distribution $D$ is used or by gathering more data from the feature space (such as adding the adversarial samples back into training data). We present such an example in Appendix~\ref{ap:ex}. Our definition of $\alpha$-effective attack leads to an impossibility of defense result that rules out any defense for any data distribution $D$ and any defense measures based on gathering more data; further, later the same definition reveals the fundamental importance of the complexity of the hypothesis class used in adversarial settings.

\subsubsec{No free lunch} In our result below, we reveal that higher the probability that an adversary is present leads to higher expected loss for any learner. In fact, we prove a stronger result: we show the above statement holds even if the learner is allowed to combine outputs in a randomized manner. Thus, we define the set of hypothesis that can be formed by choosing a set of hypothesis and randomly choosing one hypothesis: let $h_q(\mathcal{S}) = h$ such that $h \in \mathcal{S}$ and $h$ is chosen from $\mathcal{S}$ according to distribution $q$, then $R(\mathcal{H}) = \cup_{\mathcal{S} \subseteq \mathcal{H}, \mathcal{S}\mbox{ finite }} \cup_{q} \{ h_q(\mathcal{S}) \}$.

\begin{theorem} \label{nofreelunch}
Fix any hypothesis (function) class $\mathcal{H}$ and distribution $D$, and assume that an attacker exists with probability $q$. Given the attacker uses an $\alpha$-effective attack against $\mathcal{H}$ and $D$ with $\alpha \geq \alpha_0 > 0$, 
for all hypothesis $h \in R(\mathcal{H})$ the learner's loss is at least $$
E_{x,y \sim D}[l_{h^*}(x,y)] + q \alpha_0
$$
\end{theorem}

This theorem is proved in Appendix B. While the above result is a lower bound negative result, next, we present a positive upper bound result that shows that increasing the complexity of the hypothesis class considered by the learner can lead to better protection against adversarial manipulation. Towards that end, we begin by defining the lowest loss possible against a given distribution $D$: $l_{D} = \min_h E_{x,y \sim D}[ l_h(x,y)]$.
 
\begin{itemize}
\item \textbf{$\beta$-rich hypothesis class:} A hypothesis class $\mathcal{H}'$ with the following properties: (1) $\mathcal{H} \subset \mathcal{H}'$ and (2) $E_{x,y \sim D}[l_{{h}'}(x,y)] \leq \min \{l_D,E_{x,y \sim D}[l_{h^*}(x,y)] - \beta \}$ for $h' \in  \arg\!\min_{h\in \mathcal{H'}} E_{x,y \sim D}[l_{{h}}(x,y)]$, $h^* \in \arg\!\min_{h\in \mathcal{H}} E_{x,y \sim D}[l_h(x,y)]$ and all $D$. 
\end{itemize} 
Intuitively, $\mathcal{H}'$ is a more complex hypothesis class that provides lower minimum loss against any possible  distribution. 

\begin{theorem}
\label{thm2}
Fix any hypothesis (function) class $\mathcal{H}$ and distribution $D$ and a $\beta$-rich hypothesis class $\mathcal{H}'$. Assume the attacker is present with probability $q$ and $l_D <\!< E_{x,y \sim D}[l_{h^*}(x,y)] - \beta$ and $l_{\overset{\boldsymbol .}{D}} <\!< E_{x,y \sim \overset{\boldsymbol .}{D}}[l_{\overset{\boldsymbol .}{h}^*}(x,y)] - \beta$. Given the attacker that uses an $\alpha$-effective against $\mathcal{H}$ and $D$ with $\alpha = \alpha_0$
and the learner uses the $\beta$-rich hypothesis class $\mathcal{H}'$, there exists a $h \in \mathcal{H}'$ such that the loss for $h$ is less than
$$
E_{x,y \sim D}[l_{h^*}(x,y)] + q\alpha_0 - \beta
$$
\end{theorem}
This theorem is proved in Appendix C. 
There are a number of subtle points in the above result that we elaborate below:
\begin{itemize}
\item The result is an upper bound result and hence requires bounding the attacker's capabilities by imposing a bound on its effectiveness $\alpha$ (compare with Theorem~\ref{nofreelunch}).
\item The attack used is effective against the less complex class $\mathcal{H}$ whereas the defender uses the more complex class $\mathcal{H}'$. Following the  lower bound in Theorem~\ref{nofreelunch}, if the attacker were to use an effective attack against the class $\mathcal{H}'$ then the defender cannot benefit from using the rich class $\mathcal{H}'$.
\end{itemize}

Standard techniques to increase the hypothesis complexity include: considering more features, using non-linear kernels in SVM and, using a neural network with more neurons. In addition, a well known general technique---ensemble methods---is to combine any classifiers to form complex hypothesis, in which combinations of classifiers are used as the final output. 

\emph{Complexity is not free.} The above results reveal that more complex models can defend against adversaries; however, an important clarification is necessary. We assumed the existence of enough data to learn the model with high fidelity; as long as it is the case, increasing complexity leads to lower bias and better accuracy. Otherwise, learning may lead to over-fitting or high variance in the model. Thus, while the above theoretical result and recent empirical work~\cite{li2014feature} suggests more complex models for defeating adversaries, in practice, availability of data may prohibit the use of this general result. 

\begin{myprop}
	Adversaries can exploit fundamental limitations of simple hypothesis classes in providing accurate predictions in sub-regions of the feature space. Such attacks can be thwarted by moving to a more complex (richer) hypothesis class, but over-fitting issues must be addressed with the more complex class.
\end{myprop}

\section{Conclusions}
\label{sec:conclusions}

The security and privacy of machine learning is an active yet nascent area.  We have explored the attack surface of systems built upon machine learning.  That analysis yields a natural structure for reasoning about their threat models, and we have placed numerous works in this framework as organized around attacks and defenses.  We formally showed that there is often a fundamental tension between security or privacy and precision of ML predictions in machine learning systems with finite capacity.  In the large, the vast body of work from the diverse scientific communities jointly paint a picture that many vulnerabilities of machine learning and the countermeasures used to defend against them are as yet unknown--but a science for understanding them is slowly emerging.

\section*{Acknowledgments}

We thank Mart\'in Abadi, Z. Berkay Celik, Ian Goodfellow, Damien Octeau, and Kunal Talwar for feedback on early versions
of this document.
Nicolas Papernot is supported
by a Google PhD Fellowship in Security. 
We also thank Megan McDaniel for taking good care of our diet before the deadline.
Research was supported in part by the Army Research Laboratory,
under Cooperative Agreement Number W911NF-13-2-0045 (ARL Cyber Security
CRA), and the Army Research Office under grant W911NF-13-1-0421. 
The views and conclusions contained in this document are those of the
authors and should not be interpreted as representing the official policies,
either expressed or implied, of the Army Research Laboratory or the U.S.
Government. The U.S.\ Government is authorized to reproduce and distribute
reprints for government purposes notwithstanding any copyright notation hereon.

\bibliographystyle{IEEEtran}


\newpage
\appendix

\section*{\ref{ap:ex}. Example of defense using additional data}
\label{ap:ex}

\begin{figure}[h]
	\centering
	\includegraphics[width=0.9\linewidth]{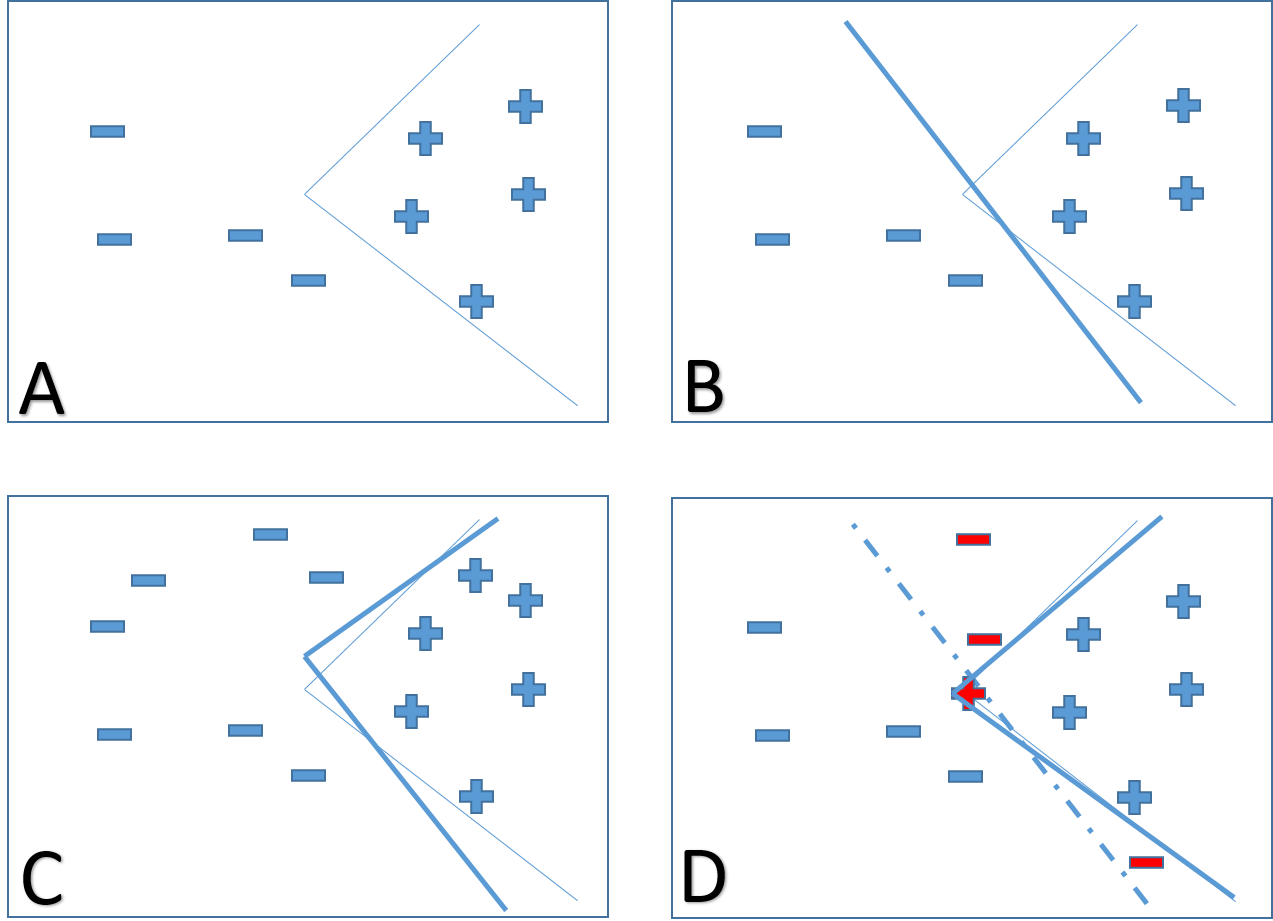}
	\caption{Defense using additional data}\label{examplefig}
\end{figure}

In Figure~\ref{examplefig} we show in sub-figure A the original data-set and the true classifier. The hypothesis class being used contains either a single linear separator or two linear separators. Thus, this hypothesis class can provide a classifier that is very close to the true classifier. However, for the data-set in A, the learned classifier is shown in sub-figure B, which is clearly far from optimal. This is not a problem of the hypothesis class; a different distribution of data shown in sub-figure C can provide for the learning of a much better classifier as shown. Another way is to add back adversarial examples as shown in sub-figure D (adversarial examples in red), which again makes the learned classifier much better.

\section*{B. Proof of Theorem~\ref{nofreelunch}}

\newtheorem*{thm:lmgfdatadep}{Theorem~\ref{nofreelunch}}
\begin{thm:lmgfdatadep}
	Fix any hypothesis (function) class $\mathcal{H}$ and distribution $D$, and assume that an attacker exists with probability $q$. Given the attacker uses an $\alpha$-effective attack against $\mathcal{H}$ and $D$ with $\alpha \geq \alpha_0 > 0$, 
	for all hypothesis $h \in R(\mathcal{H})$ the learner's loss is at least $$
	E_{x,y \sim D}[l_{h^*}(x,y)] + q \alpha_0
	$$
\end{thm:lmgfdatadep}
\begin{proof}
	Choose any hypothesis $h \in R(\mathcal{H})$. The learner's loss is $E_{x,y \sim \overset{\boldsymbol .}{D}}[l_h(x,y)] = \int l_h(x,y) \overset{\boldsymbol .}{p}(x,y) dx dy$. For a randomized classifier $h$ which randomizes over $h_1, \ldots, h_n$ with probabilities $q_1, \ldots, q_n$ respectively, the loss for any $(x,y)$ is the expected loss $\sum_i q_i l_{h_i}(x,y)$. Thus, $E_{x,y \sim \overset{\boldsymbol .}{D}}[l_h(x,y)] = \sum_i q_i \int l_{h_i}(x,y) \overset{\boldsymbol .}{p}(x,y) dx dy = \sum_i q_i E_{x,y \sim \overset{\boldsymbol .}{D}}[l_{h_i}(x,y)]$.
	Now, from the definition of $\alpha$-effective attack we have that $E_{x,y \sim \overset{\boldsymbol .}{D}}[l_{h_i}(x,y)] \geq E_{x,y \sim {D}}[l_{h^*}(x,y)] + \alpha_0$. Thus, $E_{x,y \sim \overset{\boldsymbol .}{D}}[l_h(x,y)] \geq E_{x,y \sim {D}}[l_{h^*}(x,y)] + \alpha_0$. Also, by definition $E_{x,y \sim {D}}[l_h(x,y)] \geq E_{x,y \sim {D}}[l_{h^*}(x,y)]$
	
	Next, for any choice of $h$ the adversary is present with probability $q$. Thus, the expected loss of any hypothesis is $q E_{x,y \sim \overset{\boldsymbol .}{D}}[l_h(x,y)] + (1-q) E_{x,y \sim {D}}[l_h(x,y)]$, which using the inequalities above is $\geq E_{x,y \sim D}[l_{h^*}(x,y)] + q \alpha_0$
	\end{proof}
	
\section*{C. Proof of Theorem~\ref{thm2}}
	
\newtheorem*{thm:lmgfdatadep2}{Theorem~\ref{thm2}}
\begin{thm:lmgfdatadep2}

	Fix any hypothesis (function) class $\mathcal{H}$ and distribution $D$ and a $\beta$-rich hypothesis class $\mathcal{H}'$. Assume the attacker is present with probability $q$ and $l_D <\!< E_{x,y \sim D}[l_{h^*}(x,y)] - \beta$ and $l_{\overset{\boldsymbol .}{D}} <\!< E_{x,y \sim \overset{\boldsymbol .}{D}}[l_{\overset{\boldsymbol .}{h}^*}(x,y)] - \beta$. Given the attacker that uses an $\alpha$-effective against $\mathcal{H}$ and $D$ with $\alpha = \alpha_0$
	and the learner uses the $\beta$-rich hypothesis class $\mathcal{H}'$, there exists a $h \in \mathcal{H}'$ such that the loss for $h$ is less than
	$$
	E_{x,y \sim D}[l_{h^*}(x,y)] + q\alpha_0 - \beta
	$$
\end{thm:lmgfdatadep2}
\begin{proof}
	Let the adversary's choice of distribution in his attack be $\overset{\boldsymbol .}{D}$.
	Choose the hypothesis $h \in H'$ such that $h' \in  \arg\!\min_{h\in \mathcal{H'}} E_{x,y \sim \overset{\boldsymbol .}{D}}[l_{{h}}(x,y)]$. The learner's loss is $E_{x,y \sim \overset{\boldsymbol .}{D}}[l_{h'}(x,y)]$. By definition of $\beta$-richness, we have
	$$
	E_{x,y \sim \overset{\boldsymbol .}{D}}[l_{h'}(x,y)] \leq E_{x,y \sim \overset{\boldsymbol .}{D}}[l_{\overset{\boldsymbol .}{h}^*}(x,y)] - \beta
	$$
	where $\overset{\boldsymbol .}{h}^* \in \arg\!\min_{h \in \mathcal{H}} E_{x,y \sim \overset{\boldsymbol .}{D}}[l_h(x,y)]$. Also, by definition of $\alpha$-effective attack against $\mathcal{H}$ we have
	$$
	E_{x,y \sim \overset{\boldsymbol .}{D}}[l_{\overset{\boldsymbol .}{h}^*}(x,y)] = E_{x,y \sim D}[l_{{h}^*}(x,y)] + \alpha_0
	$$
	for ${h}^* \in \arg\!\min_{h \in \mathcal{H}} E_{x,y \sim {D}}[l_h(x,y)]$. Thus, we have
	$$
	E_{x,y \sim \overset{\boldsymbol .}{D}}[l_{h'}(x,y)] \leq E_{x,y \sim D}[l_{{h}^*}(x,y)] + \alpha_0 - \beta
	$$
	Also, by $\beta$-richness we have
	$$
	E_{x,y \sim {D}}[l_{h'}(x,y)] \leq E_{x,y \sim D}[l_{{h}^*}(x,y)] - \beta
	$$
	
	Next, the adversary is present with probability $q$. Thus, the expected loss of the hypothesis $h'$ is $q E_{x,y \sim \overset{\boldsymbol .}{D}}[l_{h'}(x,y)] + (1-q) E_{x,y \sim {D}}[l_{h'}(x,y)]$, which using the inequalities above is $\leq E_{x,y \sim D}[l_{h^*}(x,y)] + q \alpha_0 - \beta$
	
\end{proof}

\newpage

\end{document}